\documentclass[prd,nofootinbib,preprint,superscriptaddress]{revtex4}
\pdfoutput=1
\usepackage[T1]{fontenc}
\usepackage{amsmath,amssymb}
\usepackage{epsfig}
\usepackage{subfigure}
\usepackage{float}
\usepackage{graphicx}
\usepackage{bigints}
\usepackage[usenames,dvipsnames]{color}
\usepackage{slashed}
\usepackage{multirow}
\usepackage[colorlinks,citecolor=blue]{hyperref}
\usepackage{pdfpages}
\usepackage{color}
\usepackage{comment}
\usepackage{ulem}
\begin{document}
	\title{Observable ${\rm \Delta{N_{eff}}}$ in Dirac Scotogenic Model}
	\author{Debasish Borah}
	\email{dborah@iitg.ac.in}
	\affiliation{Department of Physics, Indian Institute of Technology, Guwahati, Assam 781039, India}
	\author{Pritam Das}
	\email{prtmdas9@iitg.ac.in}
	\affiliation{Department of Physics, Indian Institute of Technology, Guwahati, Assam 781039, India}
	\author{Dibyendu Nanda}
	\email{dnanda@kias.re.kr}
	\affiliation{School of Physics, Korea Institute for Advanced Study, Seoul 02455, South Korea}
	\begin{abstract}
	We study the possibility of probing the radiative Dirac seesaw model with dark sector particles going inside the loop, popularly referred to as the Dirac scotogenic model via measurements of effective relativistic degrees of freedom ${\rm \Delta{N_{eff}}}$ at cosmic microwave background (CMB) experiments. The loop suppression and additional free parameters involved in neutrino mass generation allow large ($\sim\mathcal{O}(1))$ coupling of light Dirac neutrinos with the dark sector particles. Such large Yukawa coupling not only dictates the relic abundance of heavy fermion singlet dark matter but also can lead to the thermalisation of the right chiral part of Dirac neutrinos, generating additional relativistic degrees of freedom ${\rm \Delta{N_{eff}}}$. We find that the parameter space consistent with dark matter phenomenology and neutrino mass bounds can also be probed at future cosmic microwave background experiments like CMB-S4 via precision measurements of ${\rm \Delta{N_{eff}}}$. The same parameter space, while leading to loop-suppressed direct detection cross-section of dark matter outside future sensitivities, can also have other interesting and complementary observational prospects via charged lepton flavour violation and collider signatures.
	\end{abstract}

	\maketitle
	\section{Introduction}
	Evidence from astrophysics and cosmology-based observations suggests that approximately $26\%$ of our present universe is made up of dark matter (DM) \cite{Zyla:2020zbs, Planck:2018vyg}. Assuming a particle origin of DM, it is also known that none of the standard model (SM) particles can give rise to DM, leading to several beyond the standard model (BSM) proposals. Among such proposals, the weakly interacting massive particle (WIMP) paradigm has been the most well-studied one for several decades. A recent review on the status of WIMP-type DM models can be found in \cite{Arcadi:2017kky}. While the primary reason for the popularity of WIMP has been its promising detection prospects, the null results at direct detection experiments \cite{LUX-ZEPLIN:2022qhg} have pushed several simplified models to tiny corners or even motivated the particle physics community to look for alternatives beyond the WIMP paradigm. While it is premature to consider the WIMP paradigm to be ruled out, it is important to look for complementary signatures of those classes of WIMP whose direct detection prospects remain low.
	
	 The origin of neutrino mass and mixing is another motivation for pursuing BSM physics. While DM and neutrino mass could have completely different origins, it is motivating to study minimal frameworks which can relate their origins. The scotogenic model \cite{Ma:2006km} provides such an arena where non-zero neutrino masses arise at a one-loop level with dark sector particles participating in the loop. While the conventional neutrino mass models, including the original scotogenic model, predict Majorana light neutrinos, there is no experimental evidence of light neutrinos being of Majorana type. This has also led to some interest in the origin of light Dirac neutrino mass of sub-eV scale. Assuming a global unbroken lepton number symmetry, this can be minimally achieved in the SM just by incorporating three gauge singlet right-handed (RH) neutrinos, which couple to the left-handed neutrinos via the SM Higgs doublet with a tiny Yukawa coupling of the order $\lesssim 10^{-12}$. A possible extension of this minimal setup can naturally explain the smallness of light Dirac neutrino mass while also solving the puzzle of DM at the same time. In this work, we consider such a WIMP DM scenario which is not only connected to the origin of light neutrino masses, another observed phenomenon that the SM fails to address, but can also have complementary cosmological probes. In particular, we consider light neutrinos to be of Dirac nature whose sub-eV mass arises at one-loop, in a way similar to the scotogenic mechanism \cite{Ma:2006km} originally put forward for Majorana light neutrinos. The heavy scalar and fermionic fields going inside the loop are stabilized by an unbroken $Z_2$ symmetry such that the lightest $Z_2$-odd particle can be a DM candidate. We consider the extension of this model to light Dirac neutrinos. If $Z_2$-odd fermion singlet is considered to be WIMP DM, it will thermalize via Yukawa interactions leading to the thermalization of light Dirac neutrinos simultaneously. Since Dirac neutrino also contains the right chiral part, its thermalization can lead to additional relativistic degrees of freedom or dark radiation, which can be probed in cosmic microwave background (CMB) experiments. Existing CMB constraints put limits on effective
degrees of freedom for neutrinos during the
era of recombination ($z\sim 1100$) as  \cite{Planck:2018vyg} \begin{eqnarray}
{\rm
N_{eff}= 2.99^{+0.34}_{-0.33}
}
\label{Neff}
\end{eqnarray}
at $2\sigma$ or $95\%$ CL including baryon acoustic oscillation (BAO) data, which becomes more stringent to ${\rm N}_{\rm eff} = 2.99 \pm 0.17$ at $1\sigma$ CL. A similar bound also exists from big bang nucleosynthesis (BBN) $2.3 < {\rm N}_{\rm eff} <3.4$ at $95\%$ CL \cite{Cyburt:2015mya}. While these bounds are consistent with SM predictions $N^{\rm SM}_{\rm eff}=3.045$ \cite{Mangano:2005cc, Grohs:2015tfy,deSalas:2016ztq} (the recent calculation suggests a smaller value as $ N^{\rm SM}_{\rm eff}=3.0440 \pm 0.0002$ \cite{Froustey:2020mcq, Bennett:2020zkv}), future experiments like CMB Stage IV (CMB-S4) is  expected to reach an unprecedented sensitivity of $\Delta {\rm N}_{\rm eff}={\rm N}_{\rm eff}-{\rm N}^{\rm SM}_{\rm eff}
= \pm0.06$ \cite{Abazajian:2019eic}, taking it closer to the SM prediction. Enhancement of $\Delta {\rm N}_{\rm eff}$ in Dirac neutrino models have been studied in several recent works \cite{Abazajian:2019oqj, FileviezPerez:2019cyn, Nanda:2019nqy, Han:2020oet, Luo:2020sho, Borah:2020boy, Adshead:2020ekg, Luo:2020fdt, Mahanta:2021plx, Du:2021idh, Biswas:2021kio, Borah:2022obi, Li:2022yna, Biswas:2022fga, Biswas:2022vkq}. Though these works discuss the enhancement of ${\rm \Delta N_{eff}}$ due to light Dirac neutrinos, this is the first time such a possibility is explored and relevant constraints are imposed in the minimal Dirac scotogenic model.

The primary motivation of this work is to find a viable correlation among light Dirac neutrino mass, DM relic abundance and cosmological observables like $\Delta{N_{\rm eff}}$ while incorporating experimental bounds from colliders, lepton flavour violation (LFV) etc. While there exist a variety of radiative seesaw models \cite{Cai:2017jrq} with low seesaw scale and observable consequences at terrestrial experiments, radiative Dirac seesaw models have the complementary detection prospects in terms of $\Delta{N_{\rm eff}}$ due to the presence of additional light degrees of freedom. Moreover, the Dirac scotogenic model comes with the added advantage of explaining the origin of DM as well. This leads to the possibility of having a highly testable scenario where different observables in cosmic, energy and intensity frontiers can be dictated by the same particles and their couplings enhancing the predictivity. Starting with the most minimal version of the Dirac scotogenic model, we perform a comprehensive analysis taking all relevant constraints into account while also indicating the prospects of future experimental discovery via CMB as well as particle physics-based experiments. We show that in this minimal Dirac scotogenic model with fermion singlet DM, a large enhancement to $\Delta {\rm N}_{\rm eff}$ can be obtained with an interesting interplay of DM phenomenology, neutrino mass with tantalizing observational prospects via charged lepton flavour violation. 

This paper is organized as follows. In section \ref{sec1} we briefly discuss the most minimal version of the Dirac scotogenic model. In section \ref{sec2}, we discuss the details of our calculations related to DM observables, $\Delta {\rm N}_{\rm eff}$ as well as LFV. We present our results in section \ref{sec:result} and finally conclude in section \ref{sec3}.

	\section{Dirac Scotogenic Model}
	\label{sec1}
	Dirac scotogenic model has been discussed in several earlier works \cite{Gu:2007ug, Farzan:2012sa, Borah:2016zbd, Ma:2016mwh, Borah:2017leo, Wang:2017mcy, Ma:2019iwj, Ma:2019yfo, Ma:2019coj, Leite:2020wjl, Guo:2020qin, Bernal:2021ezl, Chowdhury:2022jde, Borah:2022phw} in different contexts and with different motivations. While some of these models have also incorporated additional gauge symmetries, we consider a minimal version requiring only the minimal fields augmented with discrete symmetries to realise the scenario. The SM is extended by three generations of right-handed neutrinos ($\nu_{Ri}$), three vector-like fermionic singlets ($N_i$), one inert scalar doublet ($\phi$) and one inert real scalar singlet ($\chi$). The relevant particle content of our model is projected in table \ref{tab1}.
	\begin{table}[h]
		\begin{tabular}{|c|cc|cccc|}
			\hline
			&$L$&$H$&$\nu_R$&$N$&$\phi$&$\chi$\\
			\hline
			$SU(2)$&2&2&1&1&2&1\\
			$U(1)_Y$&$-\frac{1}{2}$&$\frac{1}{2}$&0&0&$\frac{1}{2}$&0\\
			$Z_3$&0&0&$\omega$&$\omega$&$\omega$&0\\
			$Z_2$&+&+&+&-&-&-\\
			\hline
		\end{tabular}
		\caption{Particle content of the model with respective quantum numbers under the symmetry group.}\label{tab1}
	\end{table}
In addition to these new particles, a $Z_3\times Z_2$ symmetry, under which all the SM particles transform trivially, is imposed to forbid the tree-level Dirac neutrino mass term, the Majorana mass terms of $\nu_R, N$ and also to stabilise the DM particle ($N_1$). The presence of a global lepton number symmetry (under which newly introduced fermions and SM leptons carry lepton number 1) is assumed in order to forbid Majorana mass terms at higher loop levels. This choice of particles and symmetries leads to the following Yukawa interactions
	\begin{eqnarray}
		-\mathcal{L}_{\rm Yukawa}\supset\Big((y_{\phi})_{ij} \overline{L_i}\tilde{\phi}N_j+(y_{\chi})_{ij}\overline{\nu_R}_iN_j\chi+{\rm h.c.}\Big) + (M_N)_{ij} \bar{N_i}N_j\label{lag1}
	\end{eqnarray}
	where $i,j=1,2,3$ denotes three generations of fermions and $\tilde{\phi} =i\sigma^2\phi^*$. Without loss of generality, we assume $M_N$ to be diagonal. These Yukawa interactions will play a very crucial role in generating Dirac neutrino mass as well as DM relic abundance. The scalar potential of the model can be written as follows,
	\begin{eqnarray}
		V=&&\nonumber-\mu_H^2 H^\dagger H + \mu_{\phi}^2 \phi^\dagger \phi + \frac{1}{2}\mu_{ \chi}^2 \chi^2 +\frac{1}{2}\lambda_1 (H^\dagger H)^2 +\frac{1}{2} \lambda_2 (\phi^\dagger\phi)^2 +\frac{1}{4!} \lambda_3 \chi^4 \\&&+ 
		\lambda_4 (H^\dagger H) (\phi^\dagger\phi) +\frac{1}{2} \lambda_5 (H^\dagger H) \chi^2 +\frac{1}{2}\lambda_6 (\phi^\dagger\phi) \chi^2 + \lambda_7 (H^\dagger \phi) (\phi^\dagger H) \\&&+ 
		\mu (\phi^\dagger H+H^\dagger \phi) \chi\nonumber.
	\end{eqnarray}
	As mentioned, we kept the singlet scalar $\chi$ as real for simplicity. This potential does preserve $Z_2$ symmetry while $Z_3$ is softly broken by the last term, which gives us the freedom to choose $\mu$ to be very small. As discussed later, this soft breaking is necessary to generate light neutrino mass at one loop. The mass spectrum for the physical scalars, after electroweak symmetry breaking, can be obtained as follows:
	\begin{eqnarray}
		M_h^2=&&2\lambda_1 v^2;\\
		M_{\phi^\pm}^2=&&\mu^2_\phi+\lambda_4v^2;\\
		M_{\phi_I}^2=&&\mu_\phi^2+(\lambda_4+\lambda_7)v^2;\\
		M_{\chi,\phi_R}^2=&&\begin{pmatrix}
			\mu_\chi^2+\lambda_5v^2&\sqrt{2}\mu v\\\sqrt{2}\mu v&\mu_{\phi}^2+(\lambda_4+\lambda_7)v^2
		\end{pmatrix};
	\end{eqnarray}
	where $v$ denotes the vacuum expectation value (VEV) of the neutral component of the SM Higgs doublet $H$. The neutral component of the $Z_2$-odd scalar doublet is composed of real and complex parts ($\phi^0=\phi_R+i\phi_I$) and the real component of $\phi$ mixes with $\chi$ via a mixing angle $\theta$ as shown by the mass matrix squared above. Diagonalising this gives rise to two mass eigenstates having masses $M_{S_1}$ and $M_{S_2}$, which can be expressed as follows:
	\begin{equation}
		\begin{pmatrix}
			S_1\\S_2
		\end{pmatrix}=\begin{pmatrix}
			\cos\theta&\sin\theta\\-\sin\theta&\cos\theta
		\end{pmatrix}\begin{pmatrix}
			\chi\\\phi_R
		\end{pmatrix}\quad \text{where,}\quad\theta=\tan^{-1}\big[\frac{2\sqrt{2}\mu v}{\mu_{\phi}^2-\mu_{\chi}^2+(\lambda_4-\lambda_5)v^2}\big].\label{mixing}
	\end{equation}
	The relevant free parameters are the four masses ($M_{S_1},\, M_{S_2},\, M_{\phi^\pm}$ and $ M_N$), the quartic coupling ($\lambda_{5}$), mixing angle ($ \theta$) and the Yukawa couplings ($ y_{\phi},\,y_{\chi}$). They will simultaneously impact the neutrino mass generation, thermalisation of $\nu_R$, annihilation and co-annihilation rates of dark matter in the early universe. In the upcoming sections, we will discuss the importance of these parameters in detail.  
	
	\subsection{Neutrino mass and LFV}
\begin{figure}[h]
		\includegraphics[scale=0.5]{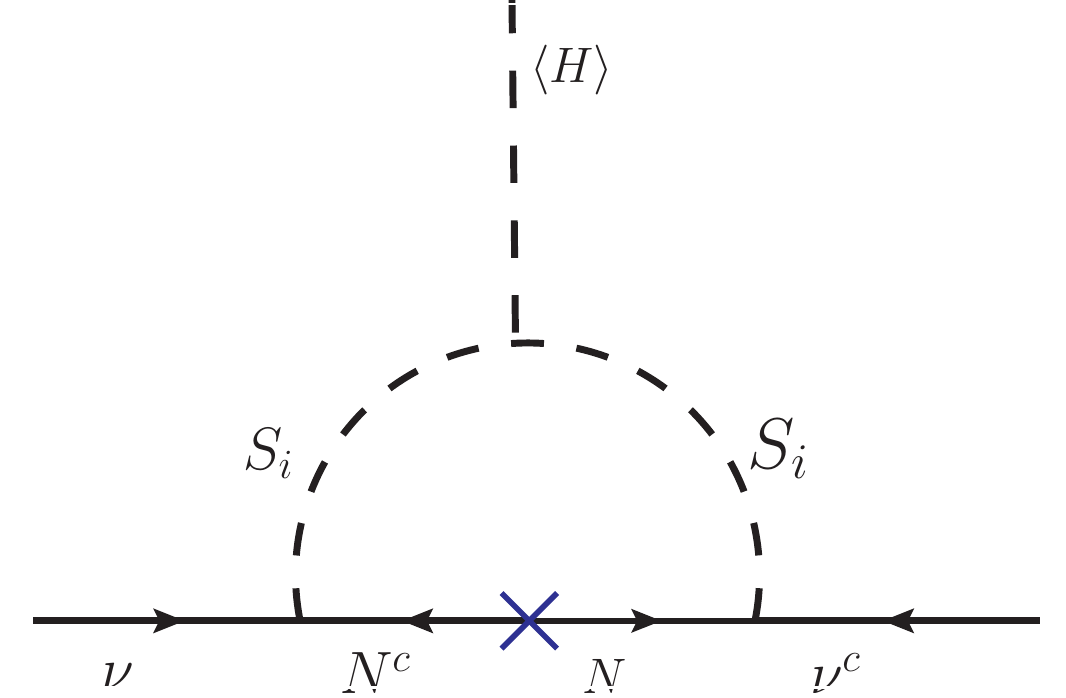}
  		\includegraphics[scale=0.5]{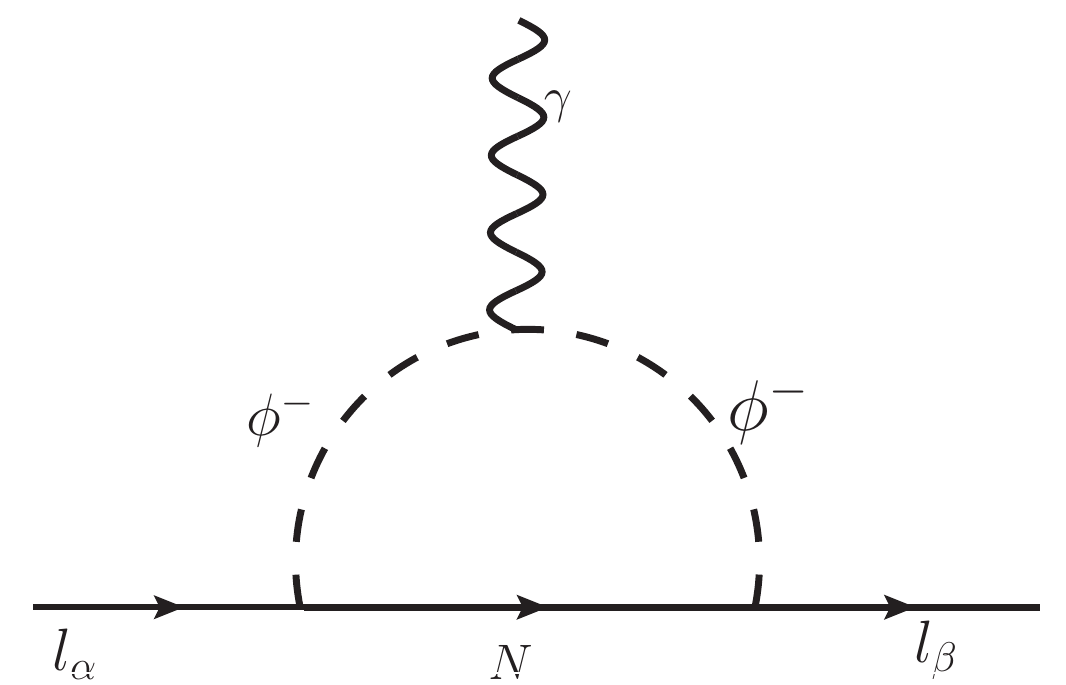}
		\caption{{\it Left panel}: Neutrino mass generation at a one-loop level in Dirac scotogenic model. Here, $S_i (i=1,2)$ is the physical mass eigenstate. {\it Right panel}: one-loop contribution charged lepton flavour violating process $l_\alpha\rightarrow l_\beta\gamma$.}\label{nmass1}
	\end{figure}

	As mentioned earlier, light Dirac neutrino mass can be generated at the one-loop level as shown on the left panel of Fig. \ref{nmass1}. The one-loop Dirac neutrino mass can be estimated as \cite{Farzan:2012sa, Guo:2020qin}:
	\begin{eqnarray}
		(M_\nu)_{\alpha\beta}=\frac{\sin2\theta}{32\pi^2\sqrt{2}}\sum_{k=1}^{3}(y_\phi)_{\alpha k}(y_\chi^*)_{\beta k}M_{N_k}\times\Big(\frac{M_{S_1}^2}{M_{S_1}^2-M_{N_k}^2}\text{ln}\frac{M_{S_1}^2}{M_{N_k}^2}-\frac{M_{S_2}^2}{M_{S_2}^2-M_{N_k}^2}\text{ln}\frac{M_{S_2}^2}{M_{N_k}^2}\Big).\label{enmass}
	\end{eqnarray} 
	Here, $M_{S_1}$ and $M_{S_2}$ are the masses of the physical scalars $S_1$ and $S_2$ which are the linear combination of $\phi_R$ and $\chi$. The present cosmological bound \cite{Planck:2018vyg} on neutrino mass suggests that the sum of absolute neutrino masses should satisfy $\sum\, m_{i}<\,0.12$ eV. Here, we have restricted ourselves to the region of parameter space where all the new particles, except $\nu_R$'s, have masses $\sim\mathcal{O} (100\, \rm GeV)$. This region of parameter space is imperative to find observable $\Delta N_{\rm eff}$ while keeping the associated Yukawa couplings within limits arising from different experimental measurements as well as perturbativity. Following Eq. \eqref{enmass}, we can get $y_\phi\,y_\chi \, \theta<10^{-7}$ to remain consistent with the neutrino mass bound. Depending on the choices of Yukawa couplings and mixing angle, which are consistent with this bound, we categorise our analysis into three separate cases, constituting the primary content of this work.
 
One interesting experimental signature of such a radiative neutrino mass model is charged lepton flavour violation. In this work, we will check the bounds from $\mu\rightarrow e\gamma$ processes only, which is tightly constrained by MEG data \cite{MEG:2016leq} while also having promising future sensitivity. It should be noted that such LFV decays depend crucially on the size of off-diagonal terms in $y_{\phi}$ and in our discussion, we have considered them equal to diagonal entries without loss of generality. Since only the scalar doublet ($\phi$) is involved with the charged lepton interactions, only the Yukawa coupling, $y_\phi$ will be responsible for the decay process. The one-loop decay can arise due to the diagram shown in the right panel of Fig. \ref{nmass1}, where $l_\alpha$ and $l_\beta$ are the two generations of charged leptons. 
The effective Lagrangian for $\mu \rightarrow e \gamma$ process is given as \cite{Kuno:1999jp}
\begin{equation}
    \mathcal{L}_{\mu \rightarrow e \gamma}=\frac{4 G_F}{\sqrt{2}}\Bigg[m_\mu A_R \bar{\mu_R} \sigma^{\mu \nu} e_L F_{\mu \nu}
    +{\rm h.c.}\Bigg],
\end{equation}
where $A_R$ is the Wilson coefficient of the dipole operator. In this model, the expression of this coefficient is calculated as \cite{Guo:2020qin}
\begin{equation}
    A_R=-\frac{\sqrt{2}}{8G_F}\frac{e(y_\phi)_{\beta i}(y^*_\phi)_{\alpha i}}{16\pi^2 M^2_{\phi^\pm}}f\big(\frac{M_{N_i}^2}{M^2_{\phi^\pm}}\big),
\end{equation}
where the loop function is defined as
\begin{equation}
    f(x)=\frac{1-6x+2x^3+3x^2(1-\text{ln}x)}{12(1-x)^2}.
\end{equation}
Finally, the decay branching ratio for $\mu\rightarrow e\gamma$ is given by:
\begin{equation}
    \text{Br} (\mu \rightarrow e \gamma)=\text{Br}(\mu \rightarrow e \nu_\mu \bar{\nu_e}) \times \frac{3\alpha_{\rm EM}}{16\pi G_F^2} \text{Abs} \Big[ \sum_i\frac{(y_\phi)_{\mu i}(y^*_{\phi})_{ei}}{M^2_{\phi^\pm}}f\big(\frac{M_{N_i}^2}{M^2_{\phi^\pm}}\big) \Big]^2.
\end{equation}
In the above equation, $G_F$ is the Fermi constant and $\alpha_{\rm EM}$ is the fine-structure constant. From the experimental point of view, the current upper limits on the $\mu\rightarrow e\gamma$ branching ratio from MEG-2016 result is Br$(\mu\rightarrow e\gamma)<4.2\times10^{-13}$ \cite{MEG:2016leq}, with future sensitivity being Br$(\mu\rightarrow e\gamma)<6\times10^{-14}$ \cite{MEGII:2018kmf}.
\subsection{Electroweak precision tests}
Due to the gauge interactions of the inert doublet $\phi$ and its mixing with $\chi$, they will contribute to $W^\pm$ and $Z$ boson self-energies. Their contributions are often parameterised in terms of the oblique parameters $S, T$ and $U$ \cite{Peskin:1990zt, Peskin:1991sw}. With the contribution to the U parameter being suppressed, the contribution to $S$ and $T$ can be written as \cite{Haber:2010bw,Beniwal:2020hjc}
 \begin{eqnarray}
 S&=& \frac{1}{\pi m_{Z}^2}\Bigg[\cos^2\theta \, \mathcal{B}_{22}\left(m_Z^2, m_{S_2}^2, m_{\phi_I}^2\right) + \sin^2\theta \, \mathcal{B}_{22}\left(m_Z^2, m_{S_1}^2, m_{\phi_I}^2\right)- \nonumber \\
 && \mathcal{B}_{22}\left(m_Z^2, m_{\phi^+}^2, m_{\phi^+}^2\right)\Bigg],\\
 T&=& \frac{1}{16\pi^2 \alpha_{\rm em}^2 v^2} \Bigg[\cos^2\theta \, \mathcal{F}\left(m_{\phi^+}^2, m_{S_2}^2 \right) + \sin^2\theta \, \mathcal{F}\left(m_{\phi^+}^2, m_{S_1}^2 \right) +\nonumber \\
 && \mathcal{F}\left(m_{\phi^+}^2, m_{\phi_I}^2\right) - \cos^2\theta \, \mathcal{F}\left(m_{S_2}^2, m_{\phi_I}^2 \right) - \sin^2\theta \, \mathcal{F}\left(m_{S_1}^2, m_{\phi_I}^2 \right)\Bigg],
 \end{eqnarray}
where, 
\begin{eqnarray}
\mathcal{B}_{22}\left(q^2, m_1^2,m_2^2 \right) &=& B_{22}\left(q^2, m_1^2, m_2^2 \right) - B_{22}\left(0, m_1^2, m_2^2 \right),\\
\mathcal{F} (x^2,y^2) &=& \frac{x^2 + y^2}{2} - \frac{x^2y^2}{x^2-y^2}\ln{\left(\frac{x^2}{y^2} \right)}.
\end{eqnarray}
$B_{22}$ is the Passarino-Veltman function \cite{Passarino:1978jh} given by
\begin{equation}
    B_{22}\left(q^2, m_1^2, m_2^2\right) = \frac{1}{4} (\Delta + 1) (m_1^2 + m_2^2 - \frac{1}{3} q^2) - \frac{1}{2} \int_{0}^1 X\ln{\left(X-i\epsilon \right)dx}
\end{equation}
where,
\begin{equation}
    X\equiv m_1^2 x + m_{2}^2(1-x) - q^2 x (1-x), \, \Delta\equiv \frac{2}{4-d} + \ln{4\pi} - \gamma_{E},
\end{equation}
with $d$ being the spacetime dimensions and $\gamma_{E}\sim 0.577$ being the Euler-Mascheroni constant. The current best-fit values of $S=0.02\pm0.07$ and $T=0.07\pm0.06$ \cite{ParticleDataGroup:2022pth} constrain the model parameter space, as we implement later in our numerical analysis.

	\subsection{Collider limits and prospects}
	Precision LEP data constrains the masses of inert scalar doublet $\phi$ components such that they do not affect the known quantities like electroweak gauge boson decay widths. For example, in order to prevent $Z$ boson decay into the neutral components of $\phi$, one must have $M_{\phi_R} + M_{\phi_I} > M_Z$. Additionally, LEP precision data also rule out the region $M_{\phi_R} < 80 \; {\rm GeV}, M_{\phi_I} < 100 \; {\rm GeV}, M_{\phi_I} - M_{\phi_R} > 8 \, {\rm GeV}$ \cite{Lundstrom:2008ai}. A similar lower bound exists on charged scalar mass $M_{\phi^\pm}> 90$ GeV. If $M_{\phi_R, \phi_I} < M_h/2$, the large hadron collider (LHC) bound on invisible Higgs decay comes into play. This, as per the recent ATLAS announcement \cite{ATLAS:2022yvh} is constrained to be below $14\%$. In terms of signatures of inert doublet, we can have either pure leptonic final states plus missing transverse energy (MET) \cite{Gustafsson:2012aj, Datta:2016nfz}, hadronic final states plus MET \cite{Poulose:2016lvz} or a mixture of both. In our model, MET may correspond to either neutrinos or singlet fermion DM. In another work \cite{Hashemi:2016wup}, tri-lepton plus MET final states were also discussed, whereas mono-jet signatures have been studied by the authors of \cite{Belyaev:2016lok, Belyaev:2018ext}. Thus, the model offers interesting collider prospects as well while also getting constrained from existing data. The details of such collider prospects is beyond the scope of this present work and can be found elsewhere. Another important observable from the collider search is the Higgs to diphoton decay rate. The presence of the charged scalar, $\phi^+$, in the inert doublet will contribute to the $ h\rightarrow \gamma \gamma$ through one loop process. The present CMS results \cite{CMS:2021kom} constrains the quantity $\frac{{\rm BR}(h\rightarrow \gamma \gamma)_{\rm expt}}{{\rm BR}(h\rightarrow \gamma \gamma)_{\rm SM}}=1.12\pm0.09$ which implies the new contribution should be in the limit,
	\begin{equation}\label{breq}
	    \frac{{\rm BR}(h\rightarrow \gamma \gamma)_{\rm New}}{{\rm BR}(h\rightarrow \gamma \gamma)_{\rm expt}}=0.03 \text{ to } 0.17.
	\end{equation}

 The additional contribution to the Higgs to diphoton branching ratio in our model arises from the charged component of the new scalar doublet ($\phi^\pm$) and the corresponding Higgs coupling ($\lambda_4$). Throughout our analysis, we keep the mass of $\phi^\pm$ within the allowed range and the Higgs portal coupling, $\lambda_4\sim\mathcal{O}(10^{-4})$, which keeps our model safe from the mentioned bound on Eq. \eqref{breq}.

	\section{Numerical Analysis}
	\label{sec2}
	
		\begin{figure}[h]
		\includegraphics[scale=0.7]{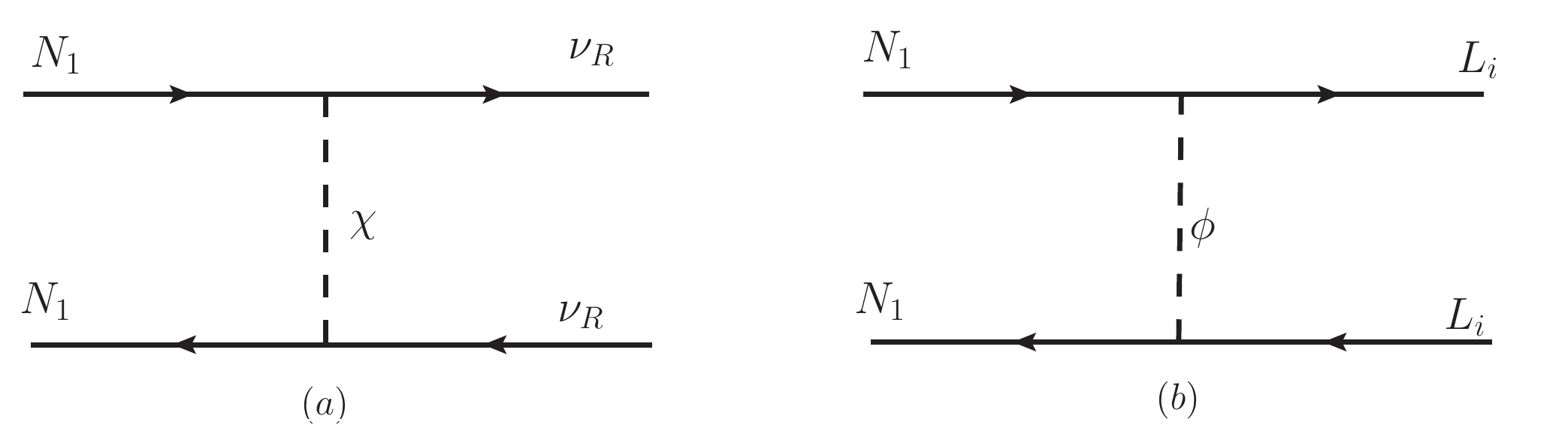}
		\caption{Dominant DM annihilation processes. In the right panel Feynman diagram, $\phi$ corresponds to all the states of $\phi$ including $\phi^0, \phi^\pm$ depending upon the final state lepton.}\label{dmanni}
	\end{figure} 
	
	\begin{figure}[h]
		\includegraphics[scale=0.55]{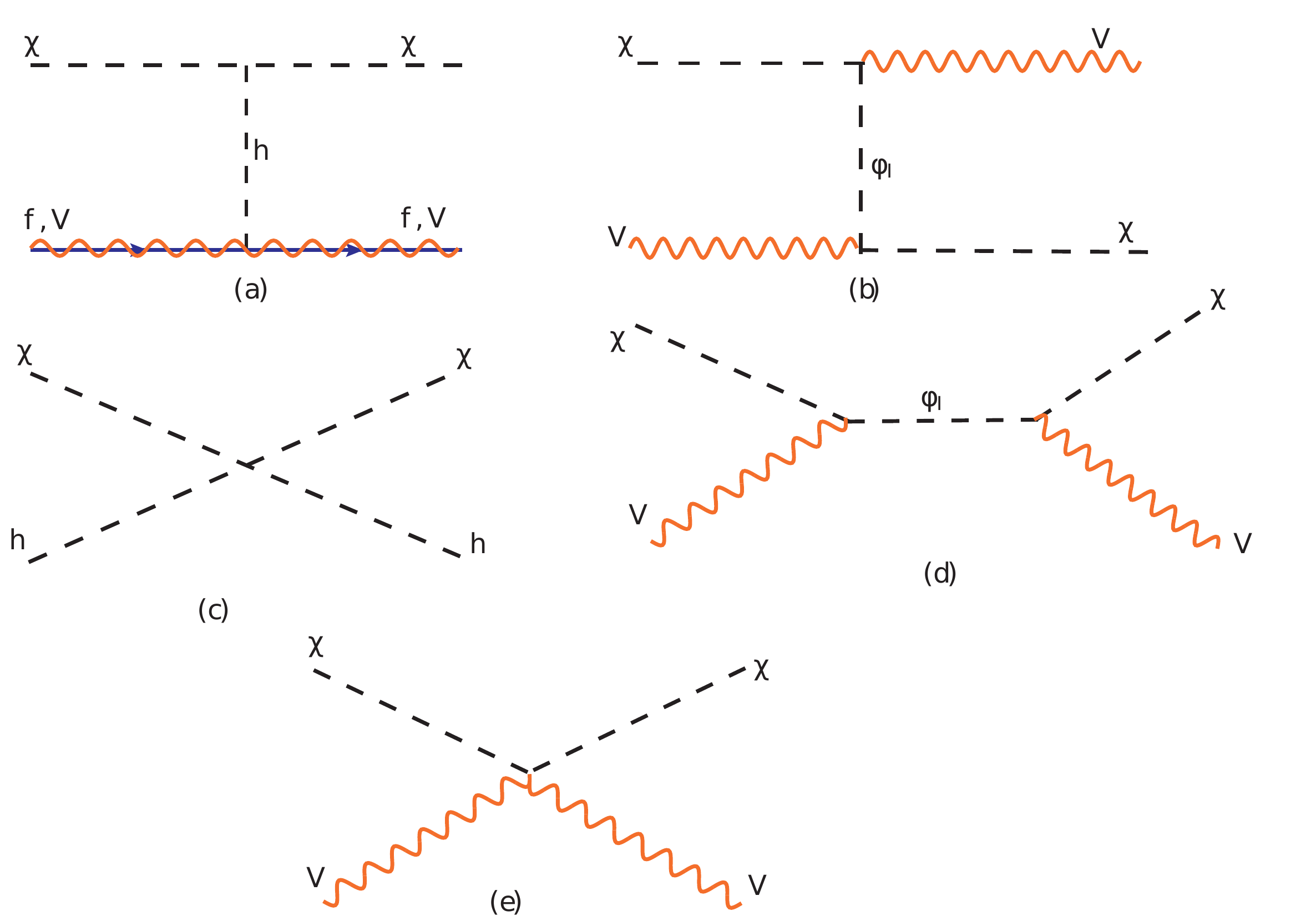}
		\caption{Scattering processes associated with thermalisation of $\chi$ with the SM bath.}\label{therm1}
	\end{figure}
	As mentioned earlier, we consider $N_1$, the lightest $Z_2$-odd particle to be the dark matter candidate. Being a gauge singlet fermion, its thermalisation and relic abundance will crucially depend upon the Yukawa interactions shown in Eq. \eqref{lag1}. The dominant annihilation channels are shown in Fig. \ref{dmanni}. However, the same Yukawa interactions are also constrained from neutrino mass criteria. However, neutrino mass also depends upon another free dimensionless parameter namely, the scalar mixing angle $\theta$. Therefore, depending upon the mixing angle and Yukawa couplings, we have categorised our study into three separate scenarios as follows.
 \begin{itemize}
     \item Case-I: In this case, the Yukawa coupling associated with the singlet scalar ($y_\chi$) is always greater than the one associated with the doublet scalar ($y_\phi$) and the mixing angle is tiny ($\sin{\theta}\leq10^{-4}$). 
     \item Case-II: In this case also the Yukawa coupling associated with the singlet scalar ($y_\chi$) is always greater than the Yukawa associated with the doublet scalar ($y_\phi$); however, the mixing angle is large ($\sin{\theta}\sim 0.7$).
     \item Case-III: In this case, both the Yukawa couplings are taken to be in the same order and the mixing angle is fixed from neutrino mass bound. 
 \end{itemize}

If the mixing angle between $\chi$ and $\phi_R$ is tiny ($\sin{\theta}\leq10^{-4}$)\footnote{When the mixing is large (case-II), we can see from Eq.\eqref{mixing} that the physical eigenstates (or the mass eigenstates) and the interaction states are different; however, if the mixing is tiny (case-I and case-III), then mass eigenstates are similar to flavour eigenstates. Therefore, we use the terminology $S_1, S_2$ for case-II only and for the other two cases, we use $\chi$ and $\phi$ explicitly.}, then the process of kinetic decoupling of $\chi$ with the SM will be influenced by processes shown by the first four diagrams of Fig. \ref{therm1}. However, in a situation, where this mixing angle is large ($\sin{\theta}\sim0.7$), the gauge boson scattering processes with $\chi$ will be the dominant channels for the kinetic decoupling process, shown by the last diagram of Fig. \ref{therm1}. The DM parameter space will not be affected in both cases since the relevant Yukawa couplings involved in DM thermalisation processes have already been fixed. On the contrary, $\Delta N_{\rm eff}$ will be affected as it is heavily dependent upon the kinetic decoupling profile of $\chi$, which acts like a portal between $\nu_R$ and the SM bath.  
 
We follow the step-by-step procedure described below to establish a connection between dark matter and effective degrees of freedom, $N_{\rm eff}$. We consider the masses for the singlet fermion and scalar as $M_{N_1}<M_\chi$ so that the decay $N_1\rightarrow\chi+\nu_R$ is kinematically forbidden. To check the viability of the lightest fermion singlet $N_1$ as a dark matter candidate, we need to check the relic abundance at the present epoch. In the first two cases, since the Yukawa coupling associated with the singlet scalar ($y_\chi$) is greater than the one with doublet scalar ($y_\phi$), the DM sector gets decoupled from the SM sector at early epochs. After the singlet scalar $\chi$ gets kinetically decoupled from the SM bath along with the fermion singlet and $\nu_R$, they do not maintain equilibrium temperature with the photons ($T_{\gamma}\ne T_{\nu_R})$. Here, we have used the notation $T_{\nu_R}$ as the temperature of the relativistic species for the dark sector temperature and $T_\gamma$ as the SM bath temperature.

  		\begin{figure}[h]
		\includegraphics[scale=0.5]{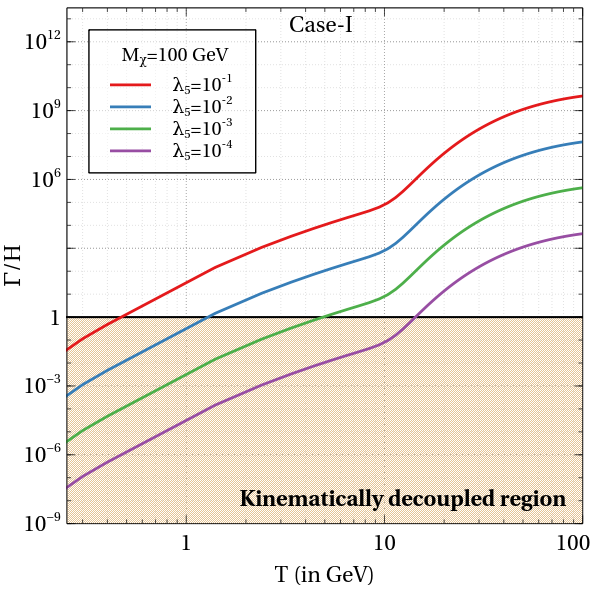}
		\includegraphics[scale=0.5]{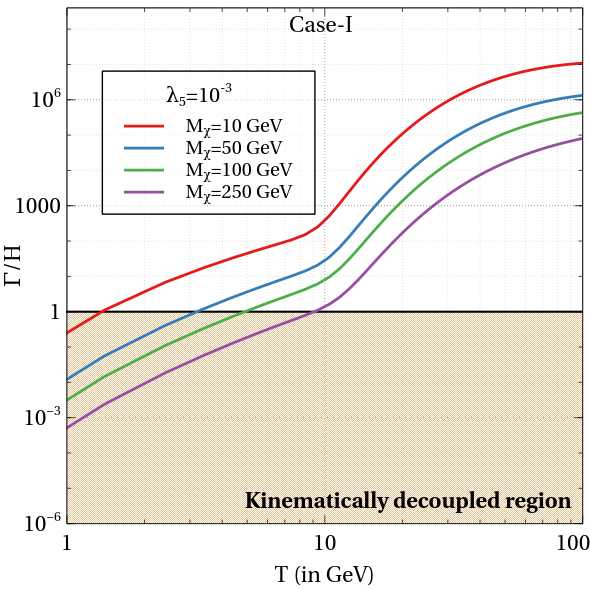}
		\includegraphics[scale=0.5]{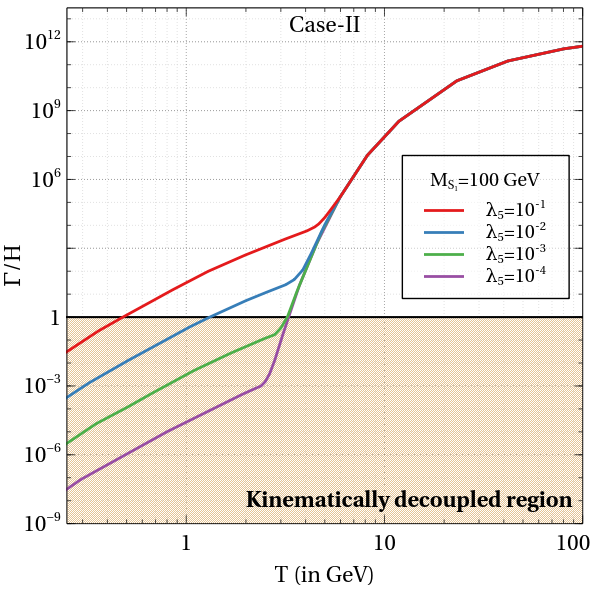}
		\includegraphics[scale=0.5]{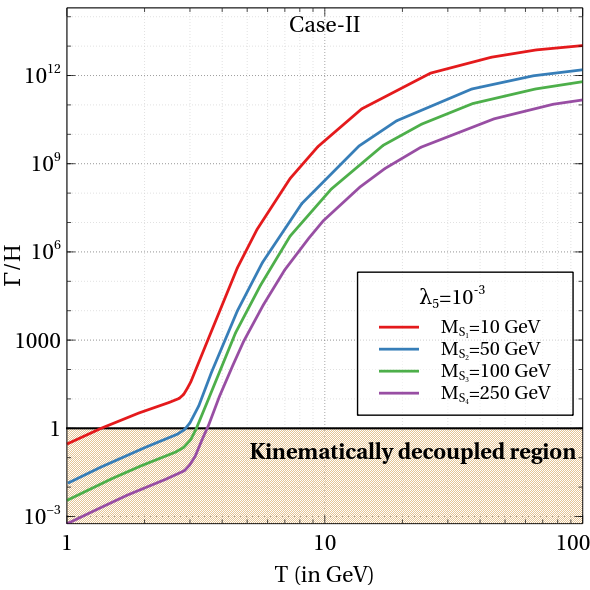}
		\caption{Comparison of $\Gamma/H$ for different choices of mass and quartic coupling $\lambda_5$ are shown for both case-I and case-II. }\label{kindec}
	\end{figure}

We first find the decoupling\footnote{As chemical decoupling has already occurred at a higher temperature, the final temperature of $\chi$ will be decided by its kinetic decoupling from the thermal bath.} temperature ($T_{\rm Dec}$) of $\chi$ for its elastic scattering processes by the simple assumption $\frac{\Gamma}{\bf H}\big \rvert_{T_{\rm Dec}}= 1$, where $\Gamma=\frac{\Gamma_{\rm el}}{n_{\rm scat}}$. Here, elastic scattering rate is $\Gamma_{\rm el}=\sum_a n_a^{\rm eq} \langle \sigma v \rangle_{\chi a\rightarrow a\chi}$ for the process $\chi a\rightarrow a\chi$,  ${\bf H}$ is the Hubble expansion rate and $a$ represents SM particles (both fermions and bosons). 
  For a massive dark matter candidate $\chi$ of mass $M_\chi$, the momentum transfer per collision with the plasma for $T \ll M_\chi$ is of the order $T$. This transferred momentum is much smaller than the average momentum of the DM candidate ($p\sim (M_\chi T)^{1/2}$).
Therefore a number of collisions $n_{scat}\sim M_\chi T$, are required for the dark matter to transfer a large part of its momentum to the plasma or to acquire it from the plasma itself \cite{Gondolo:2012vh}.However, in case-III, we adopt the trivial method to find the relic abundance \cite{Gondolo:1990dk} as in this case, the DM sector evolves together with the SM plasma due to a similar choice of Yukawa couplings ($y_\phi\sim y_\chi$) and hence they share the same bath temperature. 
We find the kinetic decoupling temperature for both the DM as well as $\nu_R$ from their respective scattering processes. Whichever a particle decouples earlier, from that decoupling temperature itself $\nu_R$ temperature will start deviating from the SM bath, simply because $\nu_R$ can interact with the SM only via dark sector particles.
	
		\begin{figure}\centering
		\includegraphics[scale=0.76]{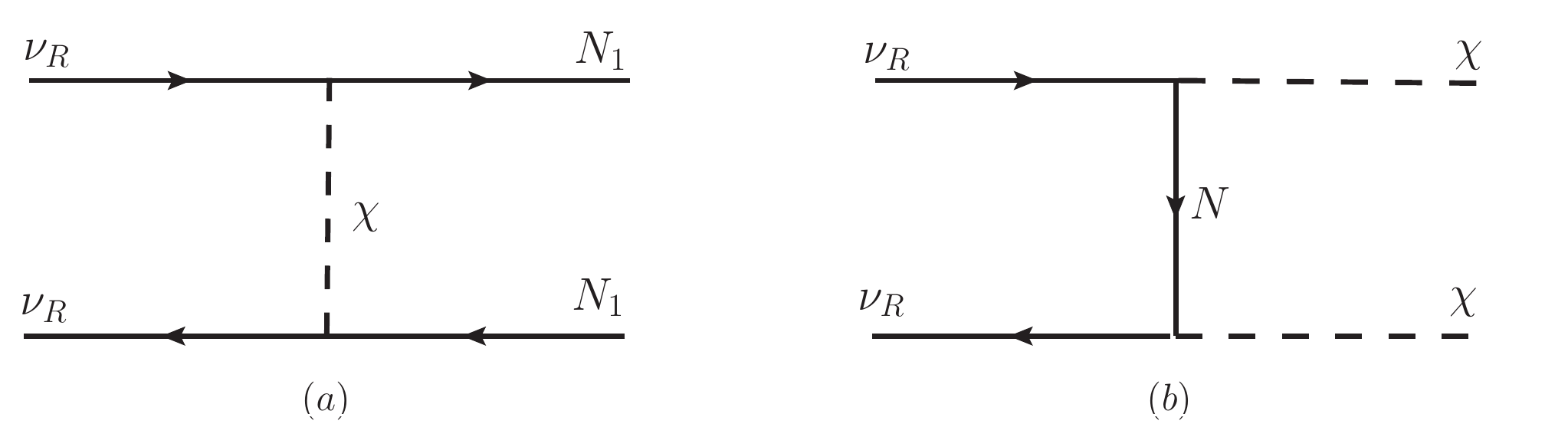}
		\caption{Thermalisation of $\nu_R$ with the dark sector.}
		\label{nuRprocess}
	\end{figure}
 
The kinetic decoupling profiles of the first two cases are shown in Fig. \ref{kindec}. In the left panel plots, we keep the singlet scalar mass fixed at 100 GeV and vary the Higgs portal quartic coupling within the range $\lambda_5 \in (10^{-1}-10^{-4})$. We have fixed the quartic coupling in the right panel plots and chosen four benchmark values for $\chi$ masses. A clear difference is visible between case-I and case-II from the left panel plots. In the lower-left panel plot, the temperature evolution patterns for the normalised scattering rate overlap at a higher temperature showing a lack of dependence on quartic coupling. This is because of the dominance of the gauge coupling over others. For small mixing angle $\theta$ in case-I, such gauge portal interactions remain ineffective, leading to changes in scattering rate for different quartic couplings even at higher temperatures, as seen from the upper left panel plot of Fig. \ref{kindec}.  The mixing depends on the Higgs VEV according to Eq. \eqref{mixing} and before the electroweak symmetry breaking (EWSB), there will not be any mixing. 
    We are interested in the situation where the dark sector decouples from the visible sector at a low temperature ($T\sim 10$ GeV) as it will lead to a significant change in the $\Delta N_{\rm eff}$. For fixed quartic coupling but the change in mass, case-II gives rise to a narrower window of decoupling temperature as compared to case-I. In case-I, for $\lambda_5$ ranging from $10^{-1}$ to $10^{-4}$, we can see decoupling temperatures shifting from 500 MeV to 10.2 GeV (top left panel plot of Fig. \ref{kindec}). On the other hand, for the singlet scalar mass ranging from 10 GeV to 250 GeV, we can see a shift in decoupling temperature from 1.2 GeV to around 10 GeV (top right panel plot of Fig. \ref{kindec}). However, in the lower panels for case-II, we can see shrinking decoupling regions for the same benchmark points due to unsuppressed gauge interactions.
The contribution to $\Delta N_{\rm eff}$ strictly depends on the kinetic decoupling temperature of $\nu_R$. From the decoupling profile study in case-II, we see a narrow separation among the decoupling temperatures as compared to case-I. This would lead to a minor relative change in $\Delta N_{\rm eff}$ calculations, hence we skip the $\Delta N_{\rm eff}$ calculation for case-II and focus completely on case-I and case-III. The thermalisation processes for $\nu_R$ are shown in Fig. \ref{nuRprocess}.

\noindent
{\bf Case-I ($y_\chi>>y_\phi)$:}  
 In this case, for the evolution of the dark sector as well as $\nu_R$, we have two distinct regions separated by the decoupling temperature $T_{\rm Dec}$: {\it Region I ($T \ge T_{\rm Dec}$}) and {\it Region II ($T\le T_{\rm Dec}$}). In region-I, all the species (SM+dark sector) are maintaining kinetic equilibrium with each other and sharing a common temperature $T_\gamma$. Therefore, we can define a reduced Boltzmann equation for the two species $\chi$ and $N_1$ (DM) into a single one with total comoving number density $Y(=\frac{n}{s})=Y_\chi+Y_{\rm DM}$ as,
	\begin{eqnarray}
		\frac{dY}{dx}=-\frac{1}{2}\frac{\beta s}{{\bf H} x} \langle \sigma v \rangle_{\rm eff}\big[Y^2-Y_{\rm eq}^2\big].\label{rel1}
	\end{eqnarray}
	Here, with an arbitrary mass scale $M_0$ we have defined $x=M_0/T$, $s(T)=g_*(T)\frac{2\pi^2T^3}{45}$ being the entropy density, ${\bf H}(T)=\sqrt{\frac{8 g_*(T)}{\pi}}\frac{T^2}{M_{\rm Pl}}$ is the Hubble rate with $g_*(T)$ as the effective relativistic degrees of freedom at temperature T. We have also defined a parameter $\beta(T)=\frac{g_*^{1/2}(T)\sqrt{g_\rho(T)}}{g_s(T)}$ with $g_s$ and $g_\rho$ being the effective relativistic degrees of freedom for entropy and energy densities respectively. Here, $g_*^{1/2}$ is defined as $g_*^{1/2}=\frac{g_s}{\sqrt{g_\rho}}\Big(1+\frac{1}{3}\frac{T}{g_s}\frac{dg_s}{dT}\Big)$. The effective annihilation cross-section for the combined processes is given by \cite{Griest:1990kh}
	\begin{eqnarray}
		\langle \sigma v \rangle_{\rm eff}=\frac{\langle \sigma v \rangle_{\rm DM\bar{DM}\rightarrow \nu_R\bar{\nu_R}}(Y_{\rm DM}^{\rm eq})^2+\langle \sigma v \rangle_{\chi\chi \rightarrow X\bar{X},\nu_R\bar{\nu_R}}(Y_\chi^{\rm eq})^2}{(Y_{\rm DM}^{\rm eq}+Y_\chi^{\rm eq})^2}.
	\end{eqnarray}
	Here, $\chi\chi \rightarrow X\bar{X}$ represents the pair annihilation of $\chi$ into the SM species through the Higgs portal interactions.
	
	Now, for region-II, the dark sector has decoupled from the SM bath and maintains a local thermodynamic equilibrium among $\chi$, $N_1$ and $\nu_R$ due to strong Yukawa interaction ($y_{\chi}$) with a common temperature $T_{\nu_R} \neq T_\gamma$. We have chosen the Yukawa $y_\phi$ to be tiny ($\sim\mathcal{O}(10^{-6})$). Such choices of parameters do satisfy the light neutrino mass and at the same time, also allow $N_i$ to evolve with the dark sector; finally, such small Yukawa coupling with the charged lepton keeps our model safe from the lepton flavour violation constraints. In this region-II, since SM and dark sector temperatures are different, we defined a quantity $\xi=\frac{T_{\nu_R}}{T_\gamma}$ to keep track of the difference in temperatures of the two sectors effectively. Therefore, we can now redefine the coupled Boltzmann equations as follows \cite{Biswas:2021kio}
	\begin{eqnarray}
		\frac{dY}{dx}=&&-\frac{1}{2}\frac{\beta s}{{\bf H} x}\langle \sigma v \rangle_{\rm eff}\big[Y^2-Y_{\rm eq}^2\big],\\
		x\frac{d\xi}{dx}+(\beta-1)\xi=&&\frac{1}{2}\frac{\beta x^4s^2}{4\alpha\xi^3{\bf H}M_0^4}\langle E\sigma v\rangle_{\rm eff}\big[Y^2-Y_{\rm eq}^2\big].\label{rel2}
	\end{eqnarray}
	Here, $\alpha=g_i\frac{7}{8}\frac{\pi^2}{30}$ with $g_i$ being the relativistic degree of freedom. We have defined the effective thermal averaged cross-section as
	\begin{eqnarray}
		\langle E\sigma v \rangle_{\rm eff}=\frac{ \langle E\sigma v\rangle'_{\nu_R\bar{\nu_R}\rightarrow \rm DM\overline{DM}}(Y_{\rm DM}^{\rm eq})^2+\langle E\sigma v\rangle'_{\nu_R\bar{\nu_R}\rightarrow \chi\chi}(Y_{\chi}^{\rm eq})^2}{(Y_{\rm DM}^{\rm eq}+Y_\chi^{\rm eq})^2},
	\end{eqnarray}
	where, $\langle E\sigma v\rangle'_{x \bar{x}\rightarrow y \bar{y}}$ is the thermal average of $E\times \sigma v_{x \bar{x}\rightarrow y\bar{y}}$ normalized by the product of equilibrium number densities of the final state particles $i.e.,~n^{\rm eq}_y n^{\rm eq}_{\bar{y}}$. The primary difference between the evolution of comoving number densities in region-I and region-II is that the former only depends on a single temperature ($T_\gamma$) while the latter's evolution is a function of both the temperatures ($T_\gamma~\&~T_{\nu_R}$). In the appendix of Ref. \cite{Biswas:2021kio}, one can find a detailed derivation of the Boltzmann equation for $\xi$ along with related terms, and cross-sections associated with it.  	
	 
	 \begin{figure}
		\includegraphics[scale=0.5]{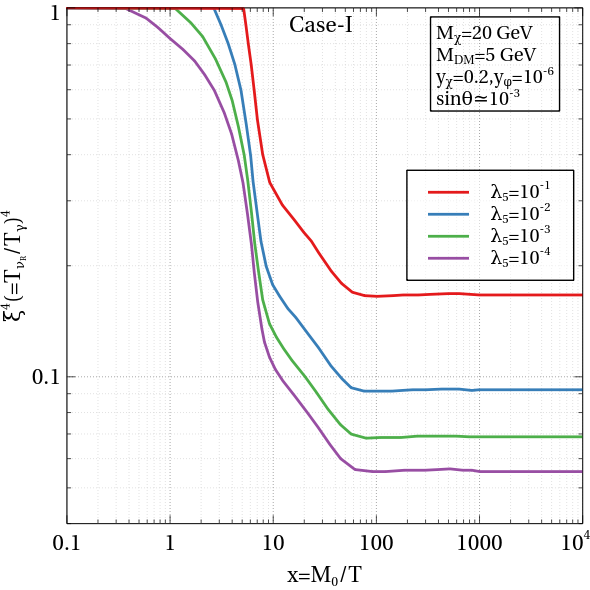}
		\includegraphics[scale=0.5]{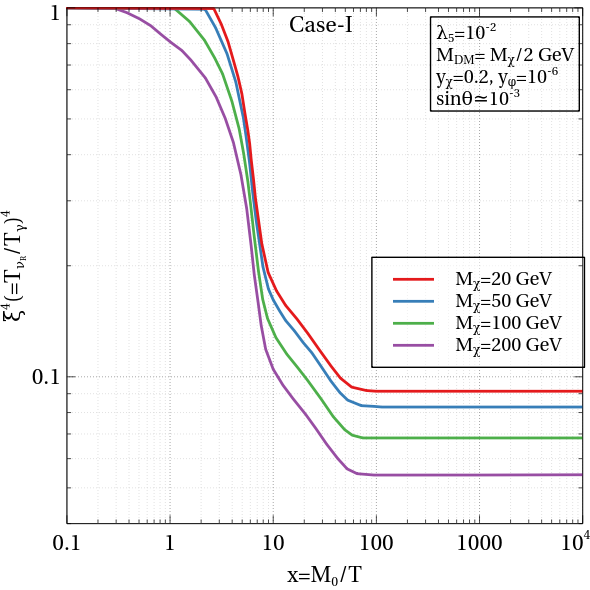}
  \includegraphics[scale=0.5]{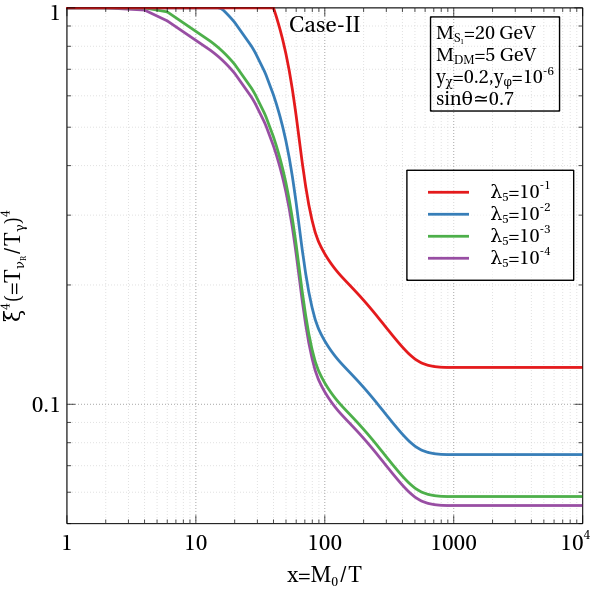}
  \includegraphics[scale=0.5]{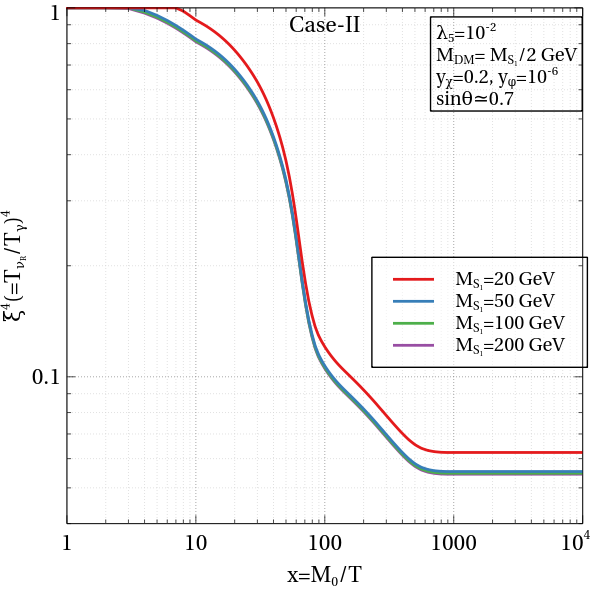}
		\caption{Variation of the temperature ratio $(\frac{T_{\nu_R}}{T_\gamma})^4$ with $x$ for different Higgs portal coupling and masses.}\label{nef2}
	\end{figure}
 We show the impact of Higgs portal couplings and $\chi$ mass on the temperature ratio ($T_{\nu_R}/T_\gamma$) in Fig. \ref{nef2}. For the tiny scalar mixing (case-I) we can see a sizable contribution for temperature ratio, which will lead to a sizable contribution to $N_{\rm eff}$. However, for the large scalar mixing parameter (case-II), the contributions are relatively weaker for the same choice of other model parameters. Therefore, further analysis of the effective relativistic degree of freedom in case-II would be less significant compared to case-I in this model.
	 \begin{figure}
    \centering
    \includegraphics[scale=0.5]{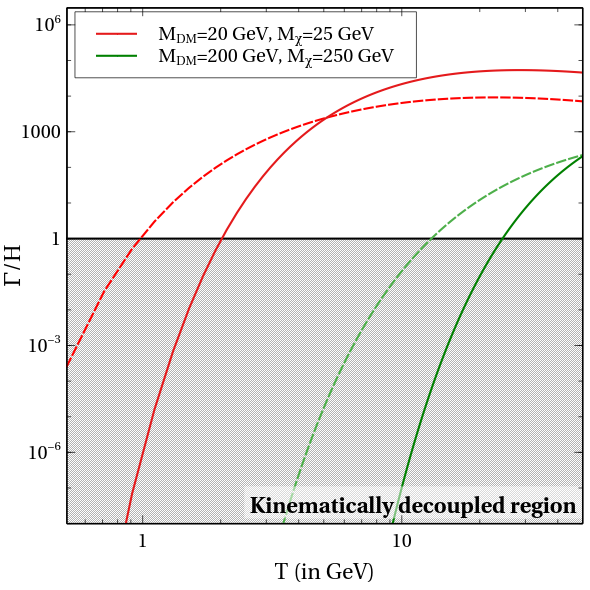}
    \includegraphics[scale=0.5]{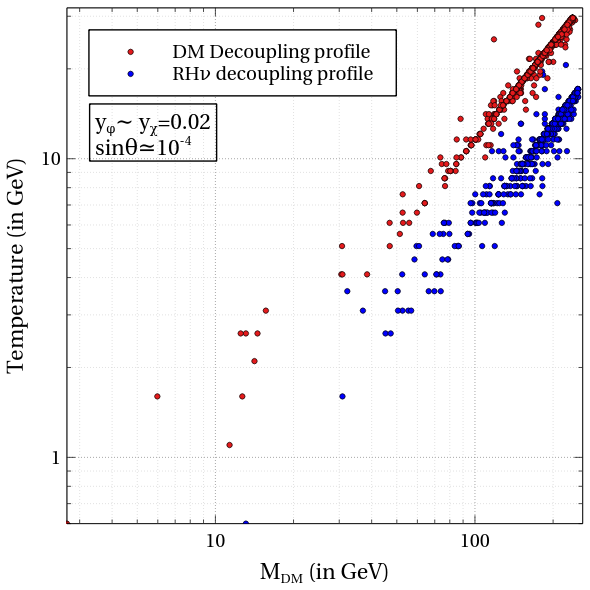}
    \caption{{\it Left panel}: The solid and dashed lines represent dark matter and $\nu_R$ decoupling patterns for two benchmark points indicated by two colours. {\it Right panel}: Decoupling temperatures of DM and right-handed neutrinos for the complete mass range.}
    \label{c3decopl}
\end{figure}
 
In case-III, due to equal strength in Yukawa couplings, both dark sector and SM evolution occur together. After dark sector freeze-out, both DM and $\nu_R$ decouple at a particular temperature. We find the kinetic decoupling temperatures for both species by comparing their scattering rates with the Hubble expansion rate as defined earlier for case-I. The comparison is shown in the left panel plot of Fig. \ref{c3decopl} while the right panel plot shows the corresponding decoupling temperatures for two benchmark points. Clearly, the DM scattering processes decouple prior to the $\nu_R$ scattering processes. 
	
	Since we are considering a Dirac neutrino framework, the right chiral counterparts $\nu_R$ are as light as the corresponding left chiral neutrinos, hence we get an additional contribution to $N_{\rm eff}$, leading to a non-zero $\Delta
	N_{\rm eff}(=N_{\rm eff}-N_{\rm eff}^{\rm SM})$. For the $N_{\rm eff}$ calculation, we have considered the instantaneous decoupling of three $\nu_R$'s and also ignored the role of lepton flavour effects. While instantaneous decoupling never happens in reality, but it can be taken as a good approximation \cite{Luo:2020sho}. As soon as the singlet scalar $\chi$ decouples from the SM plasma, it drags the dark sector along with it due to strong Yukawa interaction ($y_\chi$). We have also assumed that all the three $\nu_R$'s do have similar Yukawa coupling such that we can consider equally distributed energy density for all the three species $i.e.$, $\sum_i \rho_{\nu_{Ri}}=3\times\rho_{\nu_{Ri}}$. 
	The effective number of degrees of freedom is defined as the ratio between the radiation energy density of the non-photonic contribution to the radiation energy density of a single SM neutrino species 
 \begin{eqnarray}
     N_{\rm eff}=\frac{\rho_{\rm rad}-\rho_\gamma}{\rho_{\nu_L}}, \end{eqnarray}
where $\rho_{\nu_L}$ is the energy density of one active neutrino. Hence from this definition, we can finally define the $\Delta N_{\rm eff}$ for three generations of neutrinos as\footnote{For massless species $\rho\propto T^4$.}:
	\begin{eqnarray}	\Delta N_{\rm eff}=&&\frac{\sum_\alpha \rho_{\nu_R^\alpha}}{\rho_{\nu_L}}=3\Big(\frac{\rho_{\nu_R}}{\rho_{\nu_L}}\Big)\big \rvert_{T>T_{\nu_L}^{\rm dec}}=3\times\Big(\frac{T_{\nu_R}}{T_\gamma}\Big)^4 \big \rvert_{T>T_{\nu_L}^{\rm dec}}=3\times \xi^4.
	\end{eqnarray}
Here, we have used the bath temperature ($T_\gamma$) instead of the temperature of active neutrinos $T_{\nu_L}$. This is because, above the decoupling temperature of $\nu_L$ ($T_{\nu_L}^{\rm dec}$), both active neutrinos and photons share the same temperature. Consequently, it is not essential to track the temperature of $\nu_R$ till recombination as the ratio of $T_{\nu_R}$ and $T_{\nu_L}$ remains constant just after active neutrinos decouple from the thermal bath. Therefore, we have restricted our numerical analysis till $T_{\nu_L}^{\rm dec}$.

	\section{Results and Discussion}
	\label{sec:result}
	For numerical analysis, we considered broad parameter regions for both dark matter and $N_{\rm eff}$ calculations, as shown in table \ref{parameters} for all the cases mentioned above. As noted earlier, one of the neutral components of inert doublet can still be as light as a few GeV provided the other component masses are chosen appropriately to satisfy the LEP limits.
\begin{table}[h]
    \centering
    \begin{tabular}{|c|c||c|}
    \hline
    Cases&Parameters& Chosen range\\
    \hline
		&Dark matter mass&	$M_{\rm DM}= [1,250]$ GeV\\
	All	&Singlet scalar mass& $M_\chi= [1,350]$GeV\\
            &Doublet masses (For $\phi^\pm,\phi^0$) & $M_\phi=[1,350]$ GeV\\
            &Mass separation $(M_i-M_{\rm DM})$&$	\Delta M= [1,100]$ GeV\\
             &     Higgs portal coupling&	$\lambda_5=[10^{-4},10^{-1}]$\\
            \hline
    	Case-I&	Yukawa coupling with $\chi$ & $y_{\chi}=0.2$ ~(fixed)\\
	 \&Case-II	&Yukawa coupling with $\phi$ &~~ $y_\phi\simeq10^{-6}$ (fixed)\\
  & Mixing angle (Case-I)&$\sin{\theta}\sim10^{-6}$\\
    & Mixing angle (Case-II)&$\sin{\theta}\sim10^{-2}$\\
 \hline
&1. Yukawa coupling with $\chi~\&~\phi$ & $y_{\chi}\sim y_\phi=0.2$ ~(fixed)\\
  Case-III & Mixing angle &$\sin{\theta}\sim10^{-5}$\\
&2. Yukawa coupling with $\chi~\&~\phi$ & $y_{\chi}\sim y_\phi=0.02$ ~(fixed)\\
  & Mixing angle &$\sin{\theta}\sim10^{-4}$\\
\hline
 \end{tabular}
    \caption{The choice of parameter ranges for all three cases in our study.}
    \label{parameters}
\end{table}	
	Throughout our numerical analysis, the choices of Yukawa couplings are fixed in such a manner that they do satisfy neutrino mass bounds. We now present our results for these cases one by one.

\subsubsection{Case-I:}	For the dark matter relic calculation, we developed our own code and solved the relevant Boltzmann equations numerically to get relic abundance. As discussed earlier, the dark sector decouples from the SM sector once the kinetic equilibrium between the two sectors is lost. Therefore, we first solved the Boltzmann Eq. \eqref{rel1} till the decoupling temperature $T_{\rm Dec}$ to find the initial abundance for the dark sector particles. Then we solve the coupled BEs (Eq. \eqref{rel2}) from the $T_{\rm Dec}$ to a very low temperature to find the relic abundance of DM. Among the annihilation processes of DM, the charged scalar mediated ${\rm DM} \, \overline{\rm DM}\rightarrow l_\alpha\bar{l_\beta}$ processes remain sub-dominant due to the choice of tiny Yukawa couplings associated with charged leptons.
	\begin{figure}[h]
		\includegraphics[scale=0.5]{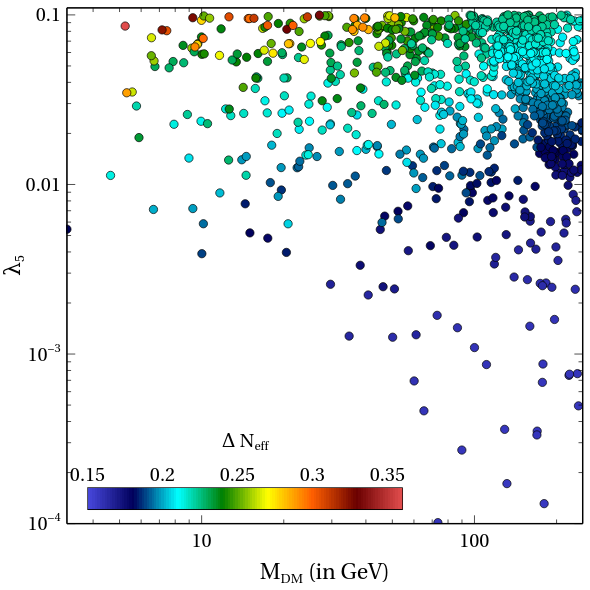}
  		\includegraphics[scale=.5]{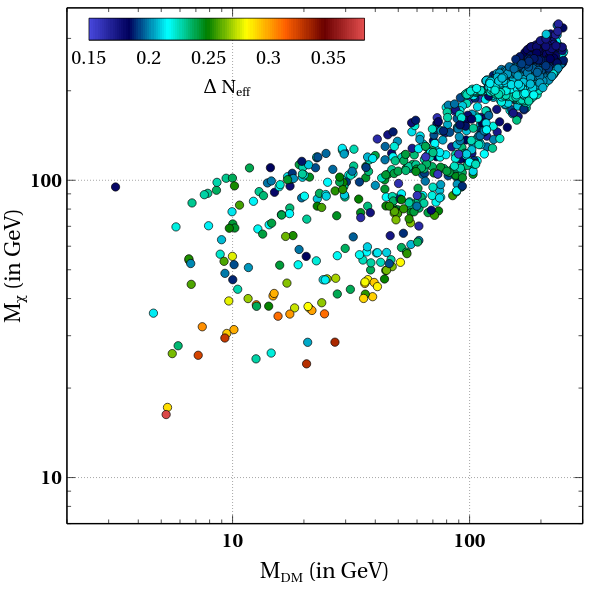}
		\caption{Allowed parameter space satisfying the correct relic density (within $3\sigma$ C.L.) of DM and all other experimental constraints such as neutrino mass, LEP and collider is shown in $M_{DM}-\lambda_{5}$ plane (Left panel) and in $M_{DM}-M_{\chi}$ plane (Right panel) where the color band represents the dependence on $\Delta{N_{\rm eff}}$.
		}
		\label{dmfig}
	\end{figure}


Fig. \ref{dmfig} shows the allowed parameter space of dark matter mass with Higgs portal couplings ($\lambda_5$ ) and the singlet scalar mass ($M_\chi$) while choosing other parameters as shown in table \ref{parameters}. We have adopted the logarithmic approach to generate random points while carrying out our numerical analysis. All the points in the plane satisfy the current $3\sigma$ bound of dark matter relic abundance ($0.117>\Omega h^2>0.123$). In the $M_{\rm DM}$ versus $M_\chi$ plot shown on the right panel, there is a clear preference for high mass DM with singlet scalar mass being close to $M_{\rm DM}$ for efficient coannihilations. From Eq. \eqref{enmass}, it is clear that both the masses, $M_{\rm DM}$ and $M_\chi$ are related to neutrino mass.

 	\begin{figure}
		\includegraphics[scale=0.5]{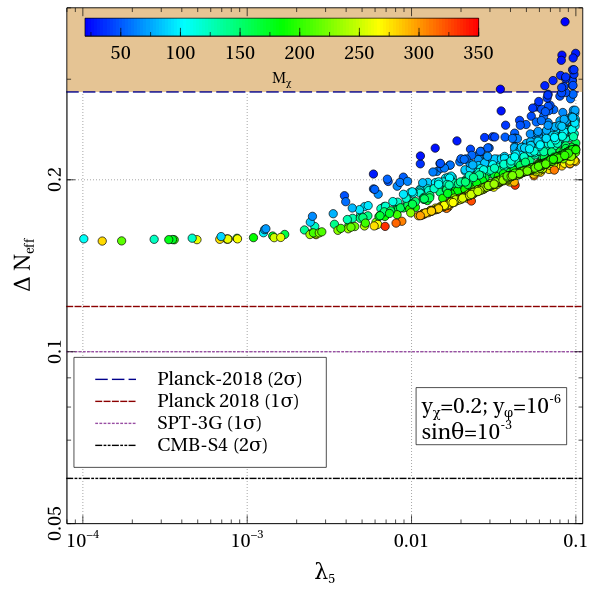}
		\includegraphics[scale=0.5]{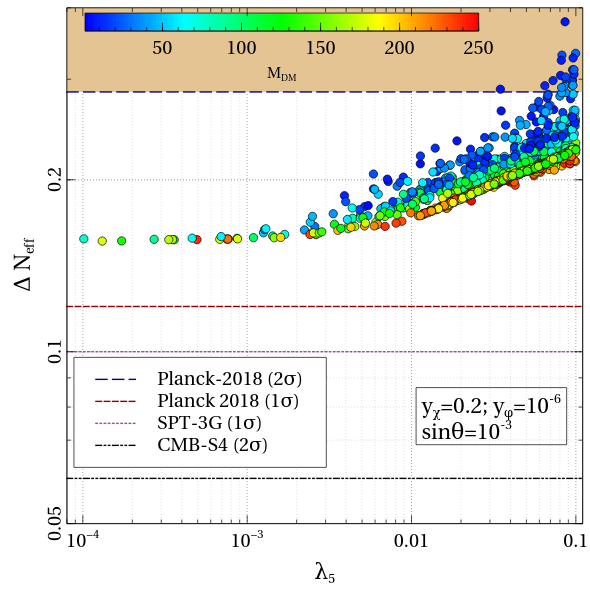}
		\caption{Dependence of $\Delta N_{\rm eff}$ with the Higgs-$\chi$ quartic coupling ($\lambda_5$) in case-I. The colour bar in the left (right) panel plot shows singlet scalar mass $M_\chi$ (dark matter mass $M_{\rm DM}$). All the points do satisfy bounds related to DM, neutrino mass, LFV, LEP and colliders.}	\label{c1neff}
	\end{figure}

In Fig. \ref{c1neff}, we show the dependence of $\Delta N_{\rm eff}$ with quartic coupling ($\lambda_5$) with the colour bars in left and right panels showing the variation in singlet scalar and DM masses respectively. One can see the gradual increase in $\Delta N_{\rm eff}$ values with an increase in $\lambda_5$ and a decrease in masses. This is expected as larger coupling or smaller masses lead to late decoupling of $\nu_R$ from the bath. The shaded region above the blue dashed horizontal line is the excluded region from the Planck 2018 constraint at $2\sigma$ C.L. ($\Delta N_{\rm eff} \leq 0.285$ \cite{Planck:2018vyg}). The other three dashed horizontal lines are respectively the bounds from Planck-2018 at $1\sigma$ C.L. ($\Delta N_{\rm eff}=0.12$ \cite{Planck:2018vyg}, red dashed line), SPT-3G at $1\sigma$ ($\Delta N_{\rm eff}=0.1$ \cite{SPT-3G:2019sok}, magenta dotted line) and CMB-S4 at $2\sigma$ ($\Delta N_{\rm eff}=0.06$ \cite{CMB-S4:2016ple}, black dot-dashed line). Clearly, the entire parameter space remains within future sensitivities.

\subsubsection{Case-III}
	In this case, we have considered two sub-scenarios by choosing two different Yukawa couplings (while maintaining equality $y_\phi=y_\chi$) with corresponding scalar mixing angles to be consistent with neutrino mass bound. The choices of numerical values of different parameters are shown in table \ref{parameters}. To carry out the numerical analysis, we first find the DM relic abundance using {\tt micrOmega} \cite{Belanger:2018ccd} and then find the decoupling profile for both the DM and $\nu_R$ for respective scattering processes. This case differs from the earlier two cases due to the choice of equal Yukawa couplings.

\begin{figure}
    \includegraphics[scale=0.5]{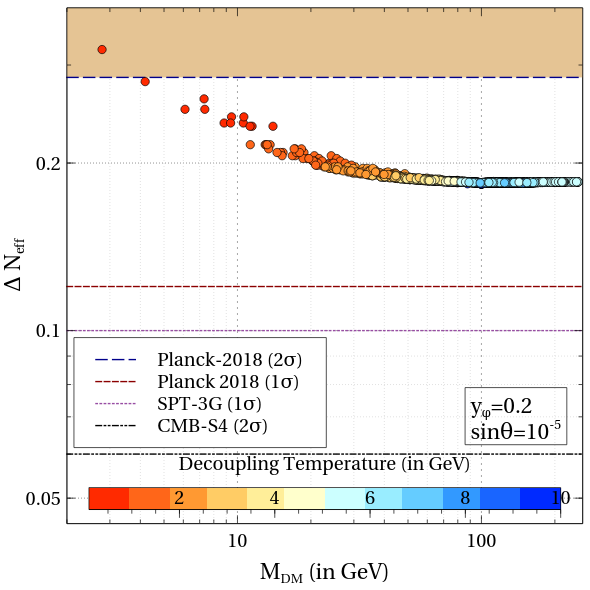}
        \includegraphics[scale=0.5]{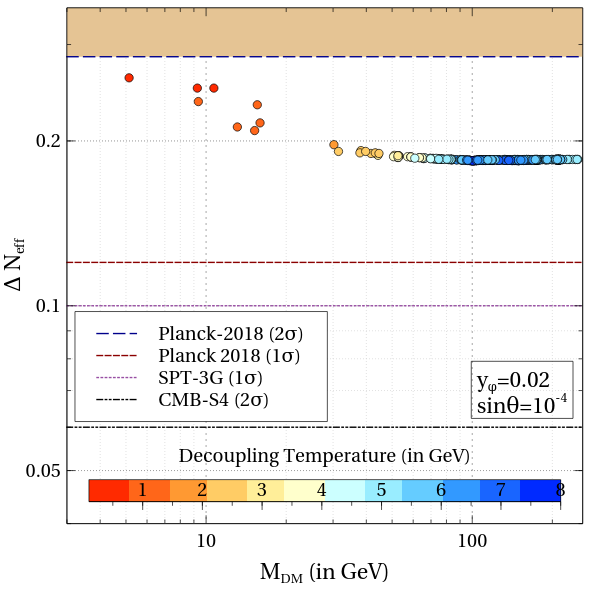}
    \caption{Dependence of $\Delta N_{\rm eff}$ with dark matter mass for case-III for two different choices of Yukawa couplings and scalar mixing angles. The colour band indicates the decoupling temperature of the dark matter candidate. All the points do satisfy bounds related to DM, neutrino mass, LEP and colliders.}
    \label{c3neff}
\end{figure}

 In Fig. \ref{c3neff}, we show how $\Delta N_{\rm eff}$ depends on dark matter mass for both the choice of Yukawa couplings. We show the decoupling temperatures in the colour band in each case. The horizontal lines correspond to the same bounds and sensitivities for $\Delta N_{\rm eff}$ as mentioned while discussing Fig. \ref{c1neff} for case-I. For most of the points, one may notice that the contribution to $\Delta N_{\rm eff}$ remains below the current Planck 2018 $2\sigma$ bound while being within future sensitivities.  

 \begin{figure}
    \centering
    \includegraphics[scale=0.6]{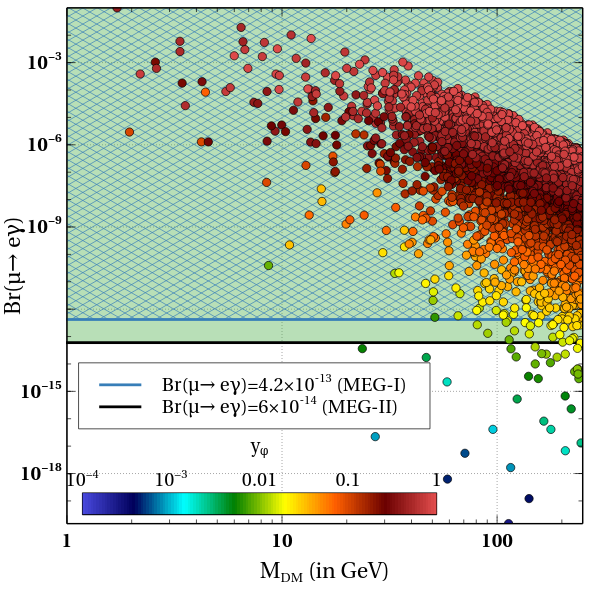}
    \caption{Br$(\mu\rightarrow e\gamma)$ versus DM mass with varying Yukawa coupling ($y_\phi$). We have shown the variation of Yukawa coupling ($y_{\phi}$) in the colour bar. The shaded regions correspond to the currently excluded regions from MEG-I \cite{MEG:2016leq} and MEG-II \cite{MEGII:2018kmf} experiments, as shown in the key.}
    \label{lfv}
\end{figure}

It should be noted that due to small Yukawa coupling $y_\phi=10^{-6}$ in case-I, case-II, it is trivial to satisfy the LFV bounds. However, this need not be so in case-III with large Yukawa couplings. Therefore, we explicitly calculate Br$(\mu\rightarrow e\gamma)$ for the parameter choices in case-III and show its variation with dark matter mass for a range of Yukawa coupling in Fig. \ref{lfv}. While in Fig. \ref{c3neff}, we considered two different choices of $y_\phi$ ($ y_\phi=0.2$ and $ y_\phi=0.02$) to show ${\rm \Delta N_{eff}}$ results,  from Fig. \ref{lfv}, we can now see that one of these choices namely, $y_\phi=0.2$ is completely ruled out from current LFV bounds. Thus, a large part of the parameter space with sizeable Yukawa coupling $y_\phi$ is already disfavoured from LFV bounds, with a considerable amount of parameter space being within reach of future LFV searches. While a future discovery of LFV decay $\mu \rightarrow e \gamma$ will definitely indicate an observable ${\rm \Delta N_{eff}}$ at future CMB experiments within the purview of this model, but not vice versa. This nevertheless keeps the complementary detection prospects of the model very promising.

\begin{figure}
    \centering
    \includegraphics[scale=0.5]{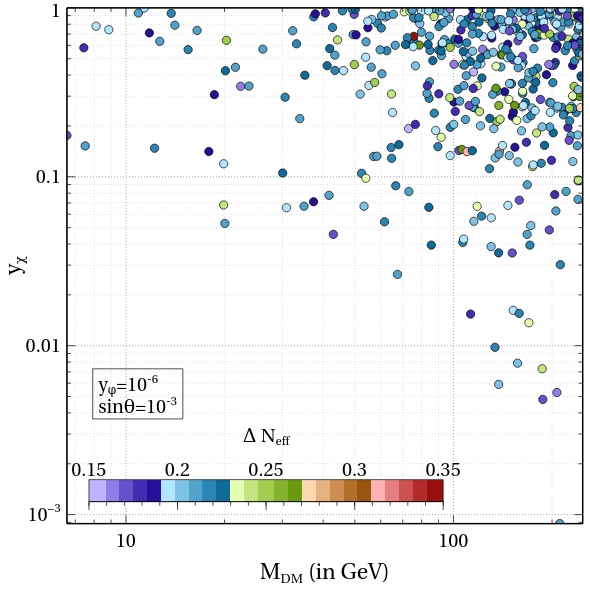}
     \includegraphics[scale=0.5]{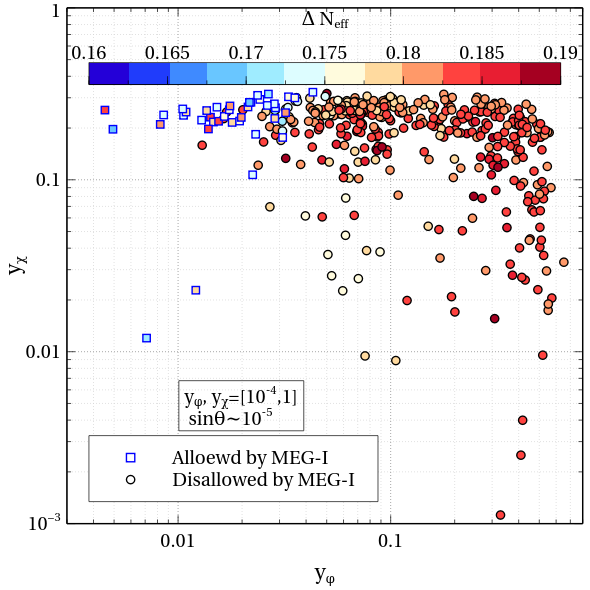}

    \caption{{\it Left panel}: Allowed parameter space in the $y_\chi-M_{DM}$ ($left$) and $y_\chi-y_\phi$ plane satisfying the current 3$\sigma$ bound on DM relic abundances, neutrino mass, LEP and collider constraints. The colour bar indicates the corresponding $\Delta N_{\rm eff}$ values.}
    \label{yukv}
\end{figure}

  \begin{figure}
    \includegraphics[scale=0.5]{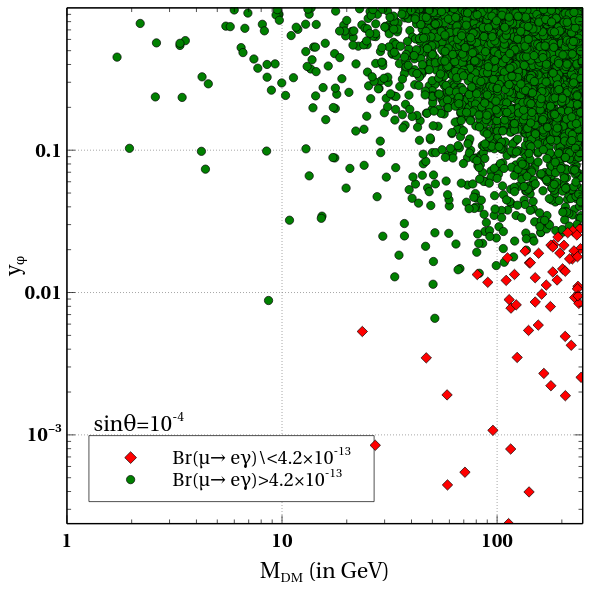}
    \includegraphics[scale=0.5]{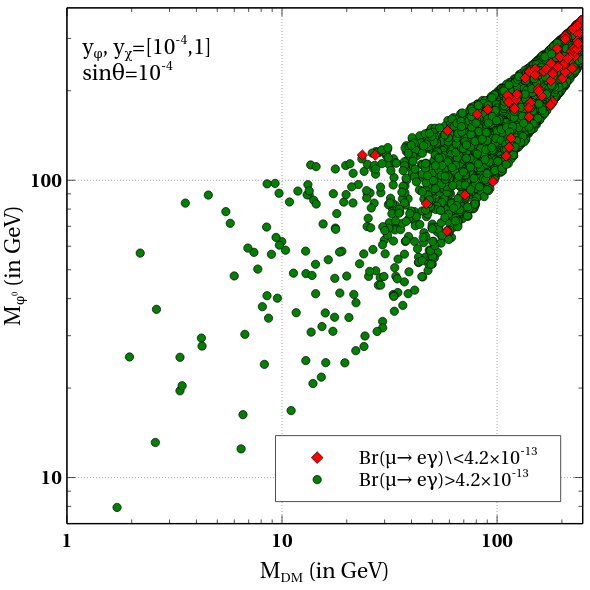}
\caption{ Allowed parameter space in the $y_\phi~vs.~M_{DM}$ ($left$) and $M_{\rm DM}~vs.~ M_{\phi^0}$ ($right$) plane satisfy the current 3$\sigma$ relic abundance bound for dark matter and relevant experimental constraints.}
    \label{c3dm}
\end{figure} 

Finally, instead of choosing specific Yukawa couplings as in case-I and case-III discussed above, we perform a numerical scan over them. In the left panel of Fig. \ref{yukv}, we show the allowed points of Yukawa coupling $y_\chi$ with dark matter mass, while the other Yukawa coupling $y_\phi$ is kept fixed at $10^{-6}$. In the right panel plot of Fig. \ref{yukv}, we show the mutual allowed points in a plane with both of these Yukawa couplings in respective axes. All the points in Fig. \eqref{yukv} satisfy the current 3$\sigma$ bound on DM relic abundances, neutrino mass and collider bounds. The colour bar in both plots indicates the corresponding $\Delta N_{\rm eff}$ value. Clearly, some of the points are already ruled out by Planck 2018 $2\sigma$ bound on $\Delta N_{\rm eff}$ and MEG-I bound on $\mu \rightarrow e \gamma$. In Fig. \ref{c3dm}, we show the final allowed parameter space allowed from all constraints except the LFV bounds. The red square-shaped (green circular) points indicate the ones allowed (disallowed) by MEG-I constraints on LFV decay $\mu \rightarrow e \gamma$. In the left panel, we have shown the allowed region in the  $y_\phi$ $vs.$ dark matter mass $M_{\rm DM}$ plane, while on the right panel, we display the allowed region in $M_{\phi^0}-M_{\rm DM}$ plane. Clearly, a large part of the parameter space already saturates the current LFV bounds.


 \begin{figure}
		\includegraphics[scale=0.7]{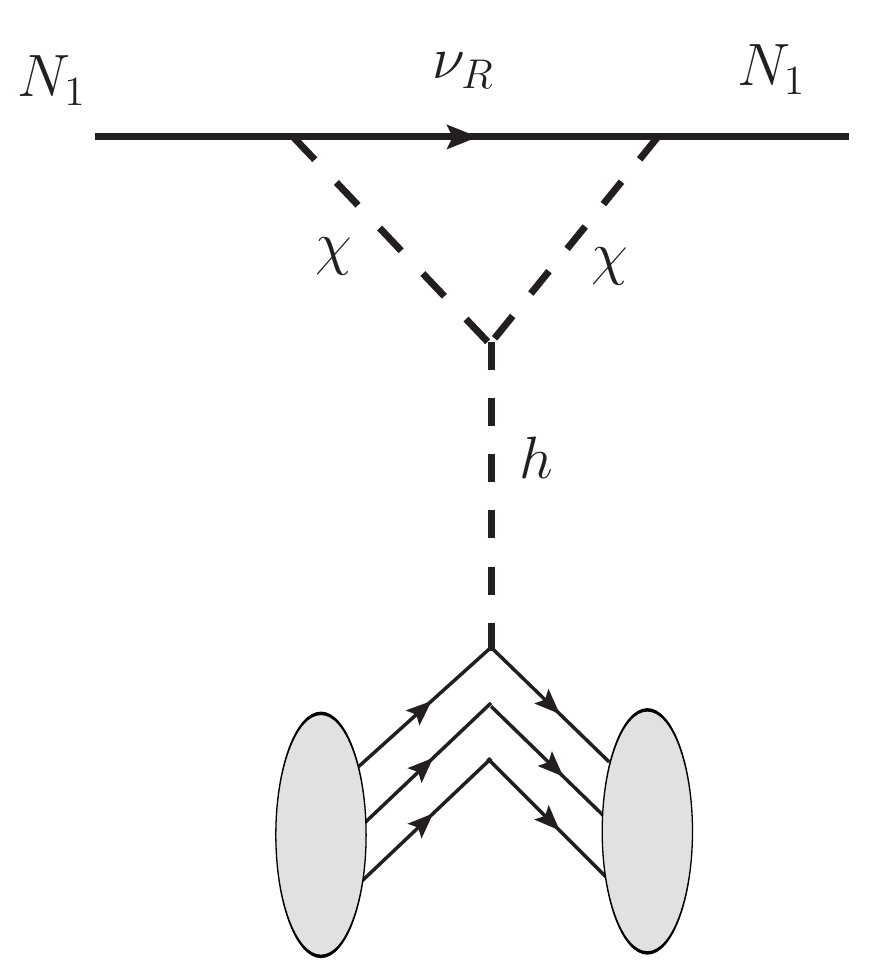}
		\caption{Schematic diagram for one-loop dark matter scattering off nucleon via SM Higgs}\label{dd1}
	\end{figure}
 \begin{figure}
      \includegraphics[scale=0.5]{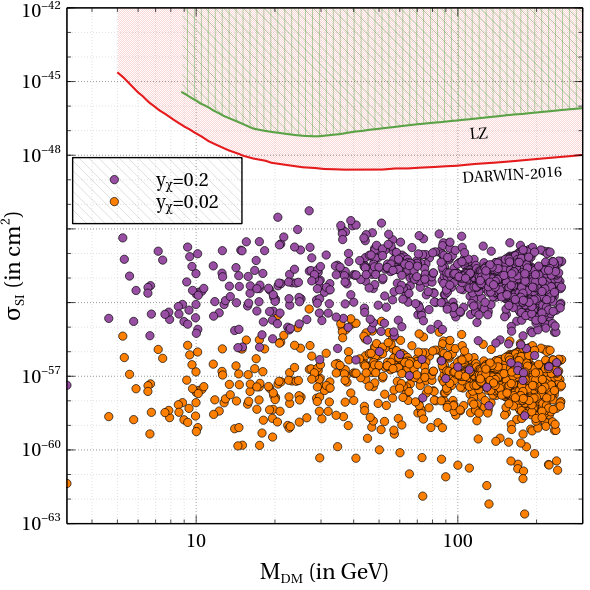}
		\caption{Spin-independent scattering cross-section of DM for two different choices of Yukawa couplings $y_\chi$.}\label{dd2}
	\end{figure}
	
Finally, we show the direct detection prospects of DM in our setup. Clearly, DM does not have any tree-level scattering process off nucleons. However, at the one-loop level, we can have a direct detection cross-section as shown in Fig. \ref{dd1}. The contribution to the spin-independent scattering cross-section for the dark matter-nucleon scattering is given by
\begin{equation}
    \sigma_{\rm SI} = \left(\frac{m_n}{v}\right)^2 \frac{\mu_{\rm DM n}^2 g_{\overline{N} N h}^2}{\pi M_{h}^4} f_{n}^2
\end{equation}
where $f_n=0.3$ depends on the quark content within a nucleon for each quark flavour\cite{Cline:2013gha}, $\mu_{\rm DM n}=\frac{M_{\rm DM} m_n}{M_{\rm DM}+m_n}$ is the reduced mass of DM-nucleon system with $m_n$ being the mass of the nucleon. The effective coupling between DM and Higgs can be written as
\begin{equation}
    g_{\overline{{N}} N h}= \frac{i}{16 \pi^2} \times \frac{y_{\chi} \lambda_{5} v}{M_{\rm DM}}\times \left[1+\left(\frac{M_{\chi}^2}{M_{\rm DM}^2} -1\right)\ln\left(1-\frac{M_{\rm DM}^2}{M_{\phi}^2} \right) \right].
\end{equation}
As only the singlet scalar $\chi$ is involved in the direct detection processes, we check the spin-independent cross-section with two choices of Yukawa couplings $y_\chi$. The other relevant parameters are kept in the same range specified in table \ref{parameters}. From Fig. \ref{dd2}, we can see that all the points in the $\sigma_{\rm SI}~vs.~ M_{\rm DM}$ plane is well below the recent bounds from LZ experiment \cite{LUX-ZEPLIN:2022qhg} shown by green curve, future sensitivity of DARWIN experiment \cite{DARWIN:2016hyl} shown by red curve\footnote{In  \cite{Biswas:2021kio}, a relatively larger value of spin-independent cross-section for the same one-loop diagram was obtained due to a missing factor in the numerical calculation.}.

 \section{Conclusion}
 \label{sec3}
We have studied the possibility of probing the Dirac scotogenic model via future CMB observations by considering fermion singlet DM. Since fermion singlet DM annihilates primarily via Yukawa interactions, the same interactions can also lead to the thermalisation of Dirac neutrinos leading to enhanced $\Delta N_{\rm eff}$. We consider all possible choices of DM Yukawa couplings as well as scalar mixing angle in our analysis. The Planck upper bound on the sum of the neutrino masses as well as the overall neutrino mass constraint, allows us to choose the Yukawa couplings and the mixing angle in such a way that gives rise to three distinct parameter regimes in this model. In each of these cases, we discuss and show the detection prospects of the model while being consistent with the desired DM phenomenology and neutrino mass constraints. While some part of the parameter space is already ruled out by the Planck 2018 upper bound on $\Delta N_{\rm eff}$, the currently allowed parameter space remains within the future sensitivity of experiments like CMB-S4. While direct detection prospects remain low for such fermion singlet DM due to radiative suppression of DM-nucleon scattering cross-section, some part of the parameter space is already ruled out by constraints from charged lepton flavour violation and collider bounds. It is worth mentioning that the detection of LFV decay $\mu\rightarrow e\gamma$ in the future would imply an observable $\Delta N_{\rm eff}$ in forthcoming CMB experiments within the framework of this Dirac scotogenic model. However, the reverse relationship is not necessarily true. These complementary detection prospects of such DM scenarios with suppressed direct detection rates are particularly interesting in view of the relentless negative searches of WIMP DM at direct detection experiments so far.  

\section{Acknowledgements}
The work of DN is supported by National Research Foundation of Korea (NRF)'s grants with grant no. 2019R1A2C3005009(DN).  PD would like to acknowledge IITG for the financial support under the project grant number: IITG/R$\&$D/IPDF/2021-22/20210911916. PD also thanks Nayan Das for the useful discussions at the early stage of this work. The work of DB is supported by Science and Engineering Research Board (SERB), Government of India grant MTR/2022/000575. 

\bibliographystyle{JHEP}
	\bibliography{Scottref, ref1, ref}

\providecommand{\href}[2]{#2}\begingroup\raggedright\begin{thebibliography}{10}

\bibitem{Zyla:2020zbs}
{\scshape Particle Data Group} collaboration, \emph{{Review of Particle
  Physics}}, \href{https://doi.org/10.1093/ptep/ptaa104}{\emph{PTEP} {\bfseries
  2020} (2020) 083C01}.

\bibitem{Planck:2018vyg}
{\scshape Planck} collaboration, \emph{{Planck 2018 results. VI. Cosmological
  parameters}},
  \href{https://doi.org/10.1051/0004-6361/201833910}{\emph{Astron. Astrophys.}
  {\bfseries 641} (2020) A6}
  [\href{https://arxiv.org/abs/1807.06209}{{\ttfamily 1807.06209}}].

\bibitem{Arcadi:2017kky}
G.~Arcadi, M.~Dutra, P.~Ghosh, M.~Lindner, Y.~Mambrini, M.~Pierre et~al.,
  \emph{{The Waning of the WIMP? A Review of Models, Searches, and
  Constraints}},  \href{https://arxiv.org/abs/1703.07364}{{\ttfamily
  1703.07364}}.

\bibitem{LUX-ZEPLIN:2022qhg}
{\scshape LUX-ZEPLIN} collaboration, \emph{{First Dark Matter Search Results
  from the LUX-ZEPLIN (LZ) Experiment}},
  \href{https://arxiv.org/abs/2207.03764}{{\ttfamily 2207.03764}}.

\bibitem{Ma:2006km}
E.~Ma, \emph{{Verifiable radiative seesaw mechanism of neutrino mass and dark
  matter}}, \href{https://doi.org/10.1103/PhysRevD.73.077301}{\emph{Phys. Rev.}
  {\bfseries D73} (2006) 077301}
  [\href{https://arxiv.org/abs/hep-ph/0601225}{{\ttfamily hep-ph/0601225}}].

\bibitem{Cyburt:2015mya}
R.~H. Cyburt, B.~D. Fields, K.~A. Olive and T.-H. Yeh, \emph{{Big Bang
  Nucleosynthesis: 2015}},
  \href{https://doi.org/10.1103/RevModPhys.88.015004}{\emph{Rev. Mod. Phys.}
  {\bfseries 88} (2016) 015004}
  [\href{https://arxiv.org/abs/1505.01076}{{\ttfamily 1505.01076}}].

\bibitem{Mangano:2005cc}
G.~Mangano, G.~Miele, S.~Pastor, T.~Pinto, O.~Pisanti and P.~D. Serpico,
  \emph{{Relic neutrino decoupling including flavor oscillations}},
  \href{https://doi.org/10.1016/j.nuclphysb.2005.09.041}{\emph{Nucl. Phys. B}
  {\bfseries 729} (2005) 221}
  [\href{https://arxiv.org/abs/hep-ph/0506164}{{\ttfamily hep-ph/0506164}}].

\bibitem{Grohs:2015tfy}
E.~Grohs, G.~M. Fuller, C.~T. Kishimoto, M.~W. Paris and A.~Vlasenko,
  \emph{{Neutrino energy transport in weak decoupling and big bang
  nucleosynthesis}},
  \href{https://doi.org/10.1103/PhysRevD.93.083522}{\emph{Phys. Rev. D}
  {\bfseries 93} (2016) 083522}
  [\href{https://arxiv.org/abs/1512.02205}{{\ttfamily 1512.02205}}].

\bibitem{deSalas:2016ztq}
P.~F. de~Salas and S.~Pastor, \emph{{Relic neutrino decoupling with flavour
  oscillations revisited}},
  \href{https://doi.org/10.1088/1475-7516/2016/07/051}{\emph{JCAP} {\bfseries
  1607} (2016) 051} [\href{https://arxiv.org/abs/1606.06986}{{\ttfamily
  1606.06986}}].

\bibitem{Froustey:2020mcq}
J.~Froustey, C.~Pitrou and M.~C. Volpe, \emph{{Neutrino decoupling including
  flavour oscillations and primordial nucleosynthesis}},
  \href{https://doi.org/10.1088/1475-7516/2020/12/015}{\emph{JCAP} {\bfseries
  12} (2020) 015} [\href{https://arxiv.org/abs/2008.01074}{{\ttfamily
  2008.01074}}].

\bibitem{Bennett:2020zkv}
J.~J. Bennett, G.~Buldgen, P.~F. De~Salas, M.~Drewes, S.~Gariazzo, S.~Pastor
  et~al., \emph{{Towards a precision calculation of $N_{\rm eff}$ in the
  Standard Model II: Neutrino decoupling in the presence of flavour
  oscillations and finite-temperature QED}},
  \href{https://doi.org/10.1088/1475-7516/2021/04/073}{\emph{JCAP} {\bfseries
  04} (2021) 073} [\href{https://arxiv.org/abs/2012.02726}{{\ttfamily
  2012.02726}}].

\bibitem{Abazajian:2019eic}
K.~Abazajian et~al., \emph{{CMB-S4 Science Case, Reference Design, and Project
  Plan}},  \href{https://arxiv.org/abs/1907.04473}{{\ttfamily 1907.04473}}.

\bibitem{Abazajian:2019oqj}
K.~N. Abazajian and J.~Heeck, \emph{{Observing Dirac neutrinos in the cosmic
  microwave background}},
  \href{https://doi.org/10.1103/PhysRevD.100.075027}{\emph{Phys. Rev.}
  {\bfseries D100} (2019) 075027}
  [\href{https://arxiv.org/abs/1908.03286}{{\ttfamily 1908.03286}}].

\bibitem{FileviezPerez:2019cyn}
P.~Fileviez~Pérez, C.~Murgui and A.~D. Plascencia, \emph{{Neutrino-Dark Matter
  Connections in Gauge Theories}},
  \href{https://doi.org/10.1103/PhysRevD.100.035041}{\emph{Phys. Rev.}
  {\bfseries D100} (2019) 035041}
  [\href{https://arxiv.org/abs/1905.06344}{{\ttfamily 1905.06344}}].

\bibitem{Nanda:2019nqy}
D.~Nanda and D.~Borah, \emph{{Connecting Light Dirac Neutrinos to a
  Multi-component Dark Matter Scenario in Gauged $B-L$ Model}},
  \href{https://arxiv.org/abs/1911.04703}{{\ttfamily 1911.04703}}.

\bibitem{Han:2020oet}
C.~Han, M.~López-Ibáñez, B.~Peng and J.~M. Yang, \emph{{Dirac dark matter in
  $U(1)_{B-L}$ with Stueckelberg mechanism}},
  \href{https://arxiv.org/abs/2001.04078}{{\ttfamily 2001.04078}}.

\bibitem{Luo:2020sho}
X.~Luo, W.~Rodejohann and X.-J. Xu, \emph{{Dirac neutrinos and $N_{{\rm
  eff}}$}}, \href{https://doi.org/10.1088/1475-7516/2020/06/058}{\emph{JCAP}
  {\bfseries 06} (2020) 058}
  [\href{https://arxiv.org/abs/2005.01629}{{\ttfamily 2005.01629}}].

\bibitem{Borah:2020boy}
D.~Borah, A.~Dasgupta, C.~Majumdar and D.~Nanda, \emph{{Observing left-right
  symmetry in the cosmic microwave background}},
  \href{https://doi.org/10.1103/PhysRevD.102.035025}{\emph{Phys. Rev. D}
  {\bfseries 102} (2020) 035025}
  [\href{https://arxiv.org/abs/2005.02343}{{\ttfamily 2005.02343}}].

\bibitem{Adshead:2020ekg}
P.~Adshead, Y.~Cui, A.~J. Long and M.~Shamma, \emph{{Unraveling the Dirac
  Neutrino with Cosmological and Terrestrial Detectors}},
  \href{https://arxiv.org/abs/2009.07852}{{\ttfamily 2009.07852}}.

\bibitem{Luo:2020fdt}
X.~Luo, W.~Rodejohann and X.-J. Xu, \emph{{Dirac neutrinos and $N_{{\rm eff}}$
  II: the freeze-in case}},  \href{https://arxiv.org/abs/2011.13059}{{\ttfamily
  2011.13059}}.

\bibitem{Mahanta:2021plx}
D.~Mahanta and D.~Borah, \emph{{Low scale Dirac leptogenesis and dark matter
  with observable $\Delta N_{\rm eff}$}},
  \href{https://arxiv.org/abs/2101.02092}{{\ttfamily 2101.02092}}.

\bibitem{Du:2021idh}
Y.~Du and J.-H. Yu, \emph{{Neutrino non-standard interactions meet precision
  measurements of $N_{\rm eff}$}},
  \href{https://arxiv.org/abs/2101.10475}{{\ttfamily 2101.10475}}.

\bibitem{Biswas:2021kio}
A.~Biswas, D.~Borah and D.~Nanda, \emph{{Light Dirac neutrino portal dark
  matter with observable \ensuremath{\Delta}Neff}},
  \href{https://doi.org/10.1088/1475-7516/2021/10/002}{\emph{JCAP} {\bfseries
  10} (2021) 002} [\href{https://arxiv.org/abs/2103.05648}{{\ttfamily
  2103.05648}}].

\bibitem{Borah:2022obi}
D.~Borah, S.~Mahapatra, D.~Nanda and N.~Sahu, \emph{{Type II Dirac Seesaw with
  Observable $\Delta N_{\rm eff}$ in the light of W-mass Anomaly}},
  \href{https://arxiv.org/abs/2204.08266}{{\ttfamily 2204.08266}}.

\bibitem{Li:2022yna}
S.-P. Li, X.-Q. Li, X.-S. Yan and Y.-D. Yang, \emph{{Effective neutrino number
  shift from keV-vacuum neutrinophilic 2HDM}},
  \href{https://arxiv.org/abs/2202.10250}{{\ttfamily 2202.10250}}.

\bibitem{Biswas:2022fga}
A.~Biswas, D.~K. Ghosh and D.~Nanda, \emph{{Concealing Dirac neutrinos from
  cosmic microwave background}},
  \href{https://doi.org/10.1088/1475-7516/2022/10/006}{\emph{JCAP} {\bfseries
  10} (2022) 006} [\href{https://arxiv.org/abs/2206.13710}{{\ttfamily
  2206.13710}}].

\bibitem{Biswas:2022vkq}
A.~Biswas, D.~Borah, N.~Das and D.~Nanda, \emph{{Freeze-in Dark Matter and
  $\Delta N_{eff}$ via Light Dirac Neutrino Portal}},
  \href{https://arxiv.org/abs/2205.01144}{{\ttfamily 2205.01144}}.

\bibitem{Cai:2017jrq}
Y.~Cai, J.~Herrero-García, M.~A. Schmidt, A.~Vicente and R.~R. Volkas,
  \emph{{From the trees to the forest: a review of radiative neutrino mass
  models}}, \href{https://doi.org/10.3389/fphy.2017.00063}{\emph{Front. in
  Phys.} {\bfseries 5} (2017) 63}
  [\href{https://arxiv.org/abs/1706.08524}{{\ttfamily 1706.08524}}].

\bibitem{Gu:2007ug}
P.-H. Gu and U.~Sarkar, \emph{{Radiative Neutrino Mass, Dark Matter and
  Leptogenesis}}, \href{https://doi.org/10.1103/PhysRevD.77.105031}{\emph{Phys.
  Rev. D} {\bfseries 77} (2008) 105031}
  [\href{https://arxiv.org/abs/0712.2933}{{\ttfamily 0712.2933}}].

\bibitem{Farzan:2012sa}
Y.~Farzan and E.~Ma, \emph{{Dirac neutrino mass generation from dark matter}},
  \href{https://doi.org/10.1103/PhysRevD.86.033007}{\emph{Phys. Rev. D}
  {\bfseries 86} (2012) 033007}
  [\href{https://arxiv.org/abs/1204.4890}{{\ttfamily 1204.4890}}].

\bibitem{Borah:2016zbd}
D.~Borah and A.~Dasgupta, \emph{{Common Origin of Neutrino Mass, Dark Matter
  and Dirac Leptogenesis}},
  \href{https://doi.org/10.1088/1475-7516/2016/12/034}{\emph{JCAP} {\bfseries
  12} (2016) 034} [\href{https://arxiv.org/abs/1608.03872}{{\ttfamily
  1608.03872}}].

\bibitem{Ma:2016mwh}
E.~Ma and O.~Popov, \emph{{Pathways to Naturally Small Dirac Neutrino Masses}},
  \href{https://doi.org/10.1016/j.physletb.2016.11.027}{\emph{Phys. Lett. B}
  {\bfseries 764} (2017) 142}
  [\href{https://arxiv.org/abs/1609.02538}{{\ttfamily 1609.02538}}].

\bibitem{Borah:2017leo}
D.~Borah and A.~Dasgupta, \emph{{Naturally Light Dirac Neutrino in Left-Right
  Symmetric Model}},
  \href{https://doi.org/10.1088/1475-7516/2017/06/003}{\emph{JCAP} {\bfseries
  06} (2017) 003} [\href{https://arxiv.org/abs/1702.02877}{{\ttfamily
  1702.02877}}].

\bibitem{Wang:2017mcy}
W.~Wang, R.~Wang, Z.-L. Han and J.-Z. Han, \emph{{The $B-L$ Scotogenic Models
  for Dirac Neutrino Masses}},
  \href{https://doi.org/10.1140/epjc/s10052-017-5446-9}{\emph{Eur. Phys. J. C}
  {\bfseries 77} (2017) 889}
  [\href{https://arxiv.org/abs/1705.00414}{{\ttfamily 1705.00414}}].

\bibitem{Ma:2019iwj}
E.~Ma, \emph{{Scotogenic cobimaximal Dirac neutrino mixing from $\Delta (27)$
  and $U(1)_\chi $}},
  \href{https://doi.org/10.1140/epjc/s10052-019-7440-x}{\emph{Eur. Phys. J. C}
  {\bfseries 79} (2019) 903}
  [\href{https://arxiv.org/abs/1905.01535}{{\ttfamily 1905.01535}}].

\bibitem{Ma:2019yfo}
E.~Ma, \emph{{Scotogenic $U(1)_\chi$ Dirac neutrinos}},
  \href{https://doi.org/10.1016/j.physletb.2019.05.006}{\emph{Phys. Lett. B}
  {\bfseries 793} (2019) 411}
  [\href{https://arxiv.org/abs/1901.09091}{{\ttfamily 1901.09091}}].

\bibitem{Ma:2019coj}
E.~Ma, \emph{{Leptonic Source of Dark Matter and Radiative Majorana or Dirac
  Neutrino Mass}},
  \href{https://doi.org/10.1016/j.physletb.2020.135736}{\emph{Phys. Lett. B}
  {\bfseries 809} (2020) 135736}
  [\href{https://arxiv.org/abs/1912.11950}{{\ttfamily 1912.11950}}].

\bibitem{Leite:2020wjl}
J.~Leite, A.~Morales, J.~W.~F. Valle and C.~A. Vaquera-Araujo,
  \emph{{Scotogenic dark matter and Dirac neutrinos from unbroken gauged B
  \ensuremath{-} L symmetry}},
  \href{https://doi.org/10.1016/j.physletb.2020.135537}{\emph{Phys. Lett. B}
  {\bfseries 807} (2020) 135537}
  [\href{https://arxiv.org/abs/2003.02950}{{\ttfamily 2003.02950}}].

\bibitem{Guo:2020qin}
S.-Y. Guo and Z.-L. Han, \emph{{Observable Signatures of Scotogenic Dirac
  Model}}, \href{https://doi.org/10.1007/JHEP12(2020)062}{\emph{JHEP}
  {\bfseries 12} (2020) 062}
  [\href{https://arxiv.org/abs/2005.08287}{{\ttfamily 2005.08287}}].

\bibitem{Bernal:2021ezl}
N.~Bernal, J.~Calle and D.~Restrepo, \emph{{Anomaly-free Abelian gauge
  symmetries with Dirac scotogenic models}},
  \href{https://doi.org/10.1103/PhysRevD.103.095032}{\emph{Phys. Rev. D}
  {\bfseries 103} (2021) 095032}
  [\href{https://arxiv.org/abs/2102.06211}{{\ttfamily 2102.06211}}].

\bibitem{Chowdhury:2022jde}
T.~A. Chowdhury, M.~Ehsanuzzaman and S.~Saad, \emph{{Dark Matter and
  $(g-2)_{\mu,e}$ in radiative Dirac neutrino mass models}},
  \href{https://arxiv.org/abs/2203.14983}{{\ttfamily 2203.14983}}.

\bibitem{Borah:2022phw}
D.~Borah, E.~Ma and D.~Nanda, \emph{{Dark SU(2) gauge symmetry and scotogenic
  Dirac neutrinos}},
  \href{https://doi.org/10.1016/j.physletb.2022.137539}{\emph{Phys. Lett. B}
  {\bfseries 835} (2022) 137539}
  [\href{https://arxiv.org/abs/2204.13205}{{\ttfamily 2204.13205}}].

\bibitem{MEG:2016leq}
{\scshape MEG} collaboration, \emph{{Search for the lepton flavour violating
  decay $\mu ^+ \rightarrow \mathrm {e}^+ \gamma $ with the full dataset of the
  MEG experiment}},
  \href{https://doi.org/10.1140/epjc/s10052-016-4271-x}{\emph{Eur. Phys. J. C}
  {\bfseries 76} (2016) 434}
  [\href{https://arxiv.org/abs/1605.05081}{{\ttfamily 1605.05081}}].

\bibitem{Kuno:1999jp}
Y.~Kuno and Y.~Okada, \emph{{Muon decay and physics beyond the standard
  model}}, \href{https://doi.org/10.1103/RevModPhys.73.151}{\emph{Rev. Mod.
  Phys.} {\bfseries 73} (2001) 151}
  [\href{https://arxiv.org/abs/hep-ph/9909265}{{\ttfamily hep-ph/9909265}}].

\bibitem{MEGII:2018kmf}
{\scshape MEG II} collaboration, \emph{{The design of the MEG II experiment}},
  \href{https://doi.org/10.1140/epjc/s10052-018-5845-6}{\emph{Eur. Phys. J. C}
  {\bfseries 78} (2018) 380}
  [\href{https://arxiv.org/abs/1801.04688}{{\ttfamily 1801.04688}}].

\bibitem{Peskin:1990zt}
M.~E. Peskin and T.~Takeuchi, \emph{{A New constraint on a strongly interacting
  Higgs sector}}, \href{https://doi.org/10.1103/PhysRevLett.65.964}{\emph{Phys.
  Rev. Lett.} {\bfseries 65} (1990) 964}.

\bibitem{Peskin:1991sw}
M.~E. Peskin and T.~Takeuchi, \emph{{Estimation of oblique electroweak
  corrections}}, \href{https://doi.org/10.1103/PhysRevD.46.381}{\emph{Phys.
  Rev.} {\bfseries D46} (1992) 381}.

\bibitem{Haber:2010bw}
H.~E. Haber and D.~O'Neil, \emph{{Basis-independent methods for the
  two-Higgs-doublet model III: The CP-conserving limit, custodial symmetry, and
  the oblique parameters S, T, U}},
  \href{https://doi.org/10.1103/PhysRevD.83.055017}{\emph{Phys. Rev. D}
  {\bfseries 83} (2011) 055017}
  [\href{https://arxiv.org/abs/1011.6188}{{\ttfamily 1011.6188}}].

\bibitem{Beniwal:2020hjc}
A.~Beniwal, J.~Herrero-Garc\'\i{}a, N.~Leerdam, M.~White and A.~G. Williams,
  \emph{{The ScotoSinglet Model: a scalar singlet extension of the Scotogenic
  Model}}, \href{https://doi.org/10.1007/JHEP06(2021)136}{\emph{JHEP}
  {\bfseries 21} (2020) 136}
  [\href{https://arxiv.org/abs/2010.05937}{{\ttfamily 2010.05937}}].

\bibitem{Passarino:1978jh}
G.~Passarino and M.~J.~G. Veltman, \emph{{One Loop Corrections for e+ e-
  Annihilation Into mu+ mu- in the Weinberg Model}},
  \href{https://doi.org/10.1016/0550-3213(79)90234-7}{\emph{Nucl. Phys. B}
  {\bfseries 160} (1979) 151}.

\bibitem{ParticleDataGroup:2022pth}
{\scshape Particle Data Group} collaboration, \emph{{Review of Particle
  Physics}}, \href{https://doi.org/10.1093/ptep/ptac097}{\emph{PTEP} {\bfseries
  2022} (2022) 083C01}.

\bibitem{Lundstrom:2008ai}
E.~Lundstrom, M.~Gustafsson and J.~Edsjo, \emph{{The Inert Doublet Model and
  LEP II Limits}},
  \href{https://doi.org/10.1103/PhysRevD.79.035013}{\emph{Phys. Rev. D}
  {\bfseries 79} (2009) 035013}
  [\href{https://arxiv.org/abs/0810.3924}{{\ttfamily 0810.3924}}].

\bibitem{ATLAS:2022yvh}
{\scshape ATLAS} collaboration, \emph{{Search for invisible Higgs-boson decays
  in events with vector-boson fusion signatures using 139 $\text{fb}^{-1}$ of
  proton-proton data recorded by the ATLAS experiment}},
  \href{https://arxiv.org/abs/2202.07953}{{\ttfamily 2202.07953}}.

\bibitem{Gustafsson:2012aj}
M.~Gustafsson, S.~Rydbeck, L.~Lopez-Honorez and E.~Lundstrom, \emph{{Status of
  the Inert Doublet Model and the Role of multileptons at the LHC}},
  \href{https://doi.org/10.1103/PhysRevD.86.075019}{\emph{Phys. Rev.}
  {\bfseries D86} (2012) 075019}
  [\href{https://arxiv.org/abs/1206.6316}{{\ttfamily 1206.6316}}].

\bibitem{Datta:2016nfz}
A.~Datta, N.~Ganguly, N.~Khan and S.~Rakshit, \emph{{Exploring collider
  signatures of the inert Higgs doublet model}},
  \href{https://doi.org/10.1103/PhysRevD.95.015017}{\emph{Phys. Rev.}
  {\bfseries D95} (2017) 015017}
  [\href{https://arxiv.org/abs/1610.00648}{{\ttfamily 1610.00648}}].

\bibitem{Poulose:2016lvz}
P.~Poulose, S.~Sahoo and K.~Sridhar, \emph{{Exploring the Inert Doublet Model
  through the dijet plus missing transverse energy channel at the LHC}},
  \href{https://doi.org/10.1016/j.physletb.2016.12.022}{\emph{Phys. Lett.}
  {\bfseries B765} (2017) 300}
  [\href{https://arxiv.org/abs/1604.03045}{{\ttfamily 1604.03045}}].

\bibitem{Hashemi:2016wup}
M.~Hashemi and S.~Najjari, \emph{{Observability of Inert Scalars at the LHC}},
  \href{https://arxiv.org/abs/1611.07827}{{\ttfamily 1611.07827}}.

\bibitem{Belyaev:2016lok}
A.~Belyaev, G.~Cacciapaglia, I.~P. Ivanov, F.~Rojas and M.~Thomas,
  \emph{{Anatomy of the Inert Two Higgs Doublet Model in the light of the LHC
  and non-LHC Dark Matter Searches}},
  \href{https://arxiv.org/abs/1612.00511}{{\ttfamily 1612.00511}}.

\bibitem{Belyaev:2018ext}
A.~Belyaev, T.~R. Fernandez Perez~Tomei, P.~G. Mercadante, C.~S. Moon,
  S.~Moretti, S.~F. Novaes et~al., \emph{{Advancing LHC probes of dark matter
  from the inert two-Higgs-doublet model with the monojet signal}},
  \href{https://doi.org/10.1103/PhysRevD.99.015011}{\emph{Phys. Rev. D}
  {\bfseries 99} (2019) 015011}
  [\href{https://arxiv.org/abs/1809.00933}{{\ttfamily 1809.00933}}].

\bibitem{CMS:2021kom}
{\scshape CMS} collaboration, \emph{{Measurements of Higgs boson production
  cross sections and couplings in the diphoton decay channel at $
  \sqrt{\mathrm{s}} $ = 13 TeV}},
  \href{https://doi.org/10.1007/JHEP07(2021)027}{\emph{JHEP} {\bfseries 07}
  (2021) 027} [\href{https://arxiv.org/abs/2103.06956}{{\ttfamily
  2103.06956}}].

\bibitem{Gondolo:2012vh}
P.~Gondolo, J.~Hisano and K.~Kadota, \emph{{The Effect of quark interactions on
  dark matter kinetic decoupling and the mass of the smallest dark halos}},
  \href{https://doi.org/10.1103/PhysRevD.86.083523}{\emph{Phys. Rev. D}
  {\bfseries 86} (2012) 083523}
  [\href{https://arxiv.org/abs/1205.1914}{{\ttfamily 1205.1914}}].

\bibitem{Gondolo:1990dk}
P.~Gondolo and G.~Gelmini, \emph{{Cosmic abundances of stable particles:
  Improved analysis}},
  \href{https://doi.org/10.1016/0550-3213(91)90438-4}{\emph{Nucl. Phys. B}
  {\bfseries 360} (1991) 145}.

\bibitem{Griest:1990kh}
K.~Griest and D.~Seckel, \emph{{Three exceptions in the calculation of relic
  abundances}}, \href{https://doi.org/10.1103/PhysRevD.43.3191}{\emph{Phys.
  Rev. D} {\bfseries 43} (1991) 3191}.

\bibitem{SPT-3G:2019sok}
{\scshape SPT-3G} collaboration, \emph{{Particle Physics with the Cosmic
  Microwave Background with SPT-3G}},
  \href{https://doi.org/10.1088/1742-6596/1468/1/012008}{\emph{J. Phys. Conf.
  Ser.} {\bfseries 1468} (2020) 012008}
  [\href{https://arxiv.org/abs/1911.08047}{{\ttfamily 1911.08047}}].

\bibitem{CMB-S4:2016ple}
{\scshape CMB-S4} collaboration, \emph{{CMB-S4 Science Book, First Edition}},
  \href{https://arxiv.org/abs/1610.02743}{{\ttfamily 1610.02743}}.

\bibitem{Belanger:2018ccd}
G.~B\'elanger, F.~Boudjema, A.~Goudelis, A.~Pukhov and B.~Zaldivar,
  \emph{{micrOMEGAs5.0 : Freeze-in}},
  \href{https://doi.org/10.1016/j.cpc.2018.04.027}{\emph{Comput. Phys. Commun.}
  {\bfseries 231} (2018) 173}
  [\href{https://arxiv.org/abs/1801.03509}{{\ttfamily 1801.03509}}].

\bibitem{Cline:2013gha}
J.~M. Cline, K.~Kainulainen, P.~Scott and C.~Weniger, \emph{{Update on scalar
  singlet dark matter}},
  \href{https://doi.org/10.1103/PhysRevD.88.055025}{\emph{Phys. Rev. D}
  {\bfseries 88} (2013) 055025}
  [\href{https://arxiv.org/abs/1306.4710}{{\ttfamily 1306.4710}}].

\bibitem{DARWIN:2016hyl}
{\scshape DARWIN} collaboration, \emph{{DARWIN: towards the ultimate dark
  matter detector}},
  \href{https://doi.org/10.1088/1475-7516/2016/11/017}{\emph{JCAP} {\bfseries
  11} (2016) 017} [\href{https://arxiv.org/abs/1606.07001}{{\ttfamily
  1606.07001}}].

\end{thebibliography}\endgroup

\end{document}